\documentclass[final,3p,times,twocolumn]{elsarticle}

\usepackage{amssymb}

\usepackage{lineno}

\usepackage{amsmath}
\usepackage{tikz}
\usepackage{array}
\usepackage[parfill]{parskip}
\usepackage{xargs} 
\usepackage{tabularx} 
\usepackage{colortbl} 
\usepackage{float} 
\usepackage{xcolor}
\usepackage{xr}
\usepackage{makecell}
\usepackage{supertabular}
\usepackage{bbm}
\usepackage[british]{babel}

\usetikzlibrary{calc,trees,positioning,arrows,chains,shapes.geometric,  decorations.pathreplacing,decorations.pathmorphing,shapes,matrix,shapes.symbols}

\emergencystretch=\maxdimen
\definecolor{C0}{rgb}{0.12156862745098,0.466666666666667,0.705882352941177}
\definecolor{C1}{rgb}{1,0.498039215686275,0.0549019607843137}
\definecolor{C2}{rgb}{0.172549019607843,0.627450980392157,0.172549019607843}
\definecolor{C3}{rgb}{0.83921568627451,0.152941176470588,0.156862745098039}
\definecolor{C4}{rgb}{0.580392156862745,0.403921568627451,0.741176470588235}
\definecolor{C5}{rgb}{0.549019607843137,0.337254901960784,0.294117647058824}
\definecolor{C6}{rgb}{0.890196078431372,0.466666666666667,0.76078431372549}
\colorlet{C7}{gray!99.607843137254903!black}
\definecolor{C8}{rgb}{0.737254901960784,0.741176470588235,0.133333333333333}
\definecolor{C9}{rgb}{0.0901960784313725,0.745098039215686,0.811764705882353}
\definecolor{lightgray}{rgb}{0.9,0.9,0.9}

\tikzstyle{decision} = [diamond, aspect=2, draw=C2, line width=1mm,
    text width=5.5em, text badly centered, node distance=2.5cm, inner sep=0pt]
\tikzstyle{block} = [rectangle, draw=C0, line width=1mm,
    text width=8em, text centered, rounded corners, minimum height=2em]
\tikzstyle{redblock} = [rectangle, draw=C3, line width=1mm,
    text width=10em, text centered, rounded corners, minimum height=2em]
\tikzstyle{redblock_wide} = [rectangle, draw=C3, line width=1mm,
    text width=16em, text centered, rounded corners, minimum height=2em]
\tikzstyle{iteration} = [rectangle, draw=C3, line width=1mm,
    text width=10em, text centered, rounded corners, minimum height=2em] 
\tikzstyle{background} = [rectangle, draw=C3, line width=1mm, minimum width=5em, rounded corners, minimum height=2em] 
\tikzstyle{arrow} = [draw, thick,->,>=stealth]

\usepackage[colorinlistoftodos,prependcaption,textsize=tiny]{todonotes}
\newcommandx{\introduction}[2][1=]{\todo[linecolor=red,backgroundcolor=red!25,bordercolor=red,#1]{#2}}
\newcommandx{\change}[2][1=]{\todo[linecolor=blue,backgroundcolor=blue!25,bordercolor=blue,#1]{#2}}
\newcommandx{\info}[2][1=]{\todo[linecolor=C2,backgroundcolor=C1!25,bordercolor=C0,#1]{#2}}
\newcommandx{\TODO}[2][1=]{\todo[linecolor=orange,backgroundcolor=orange!25,bordercolor=orange,#1]{#2}}
\newcommandx{\improvement}[2][1=]{\todo[linecolor=C4,backgroundcolor=C4!25,bordercolor=C4,#1]{#2}}
\newcommandx{\thiswillnotshow}[2][1=]{\todo[disable,#1]{#2}}
\newcommandx{\OTH}[2][1=]{\todo[linecolor=C0,backgroundcolor=C0!25,bordercolor=C0,#1]{\color{blue}\bf OTH: #2}}
\newcommandx{\LH}[2][1=]{\todo[linecolor=C3,backgroundcolor=C3!25,bordercolor=C3,#1]{\color{C3}\bf LH: #2}}
\newcommandx{\AJ}[2][1=]{\todo[linecolor=C2,backgroundcolor=C2!25,bordercolor=C2,#1]{\color{C2}\bf AJ: #2}}
\newcommandx{\ATE}[2][1=]{\todo[linecolor=C4,backgroundcolor=C4!25,bordercolor=C4,#1]{\color{C4}\bf ATE: #2}}

\newcommand{\configuration}{{configuration}}
\newcommand{\configurations}{{configurations}}
\newcommand{\Configuration}{{Configuration}}
\newcommand{\Configurations}{{Configurations}}
\newcommand{\closepacked}{close-packed~}

\newcommand{\localgeometry}{{local geometry}}
\newcommand{\localgeometries}{{local geometries}}
\newcommand{\Localgeometries}{{Local geometries}}
\newcommand{\LocalGeometries}{{Local Geometries}}
\newcommand{\abinitiothermodynamics}{ab initio thermodynamics}
\newcommand{\localadsorptiongeometry}{{symmetry-inequivalent local geometry}}
\newcommand{\localadsorptiongeometries}{{symmetry-inequivalent local geometries}}
\newcommand{\Localadsorptiongeometries}{{Symmetry-inequivalent local geometries}}

\newcommand{\StandardUnitCell}{{Standard Unit Cell}}
\newcommand{\standardunitcell}{{standard unit cell}}
\newcommand{\standardunitcells}{{standard unit cells}}
\newcommand{\configurationhash}{{configuration hash}}
\newcommand{\configurationhashes}{{configuration hashes}}
\newcommand{\FeatureVector}{{Feature Vector}}
\newcommand{\featurevector}{{feature vector}}
\newcommand{\featurevectors}{{feature vectors}}
\newcommand{\featuredimension}{{feature dimension}}
\newcommand{\featurethreshold}{{feature threshold}}
\newcommand{\likelihood}{{likelihood}}
\newcommand{\posterior}{{posterior probability}}
\newcommand{\posteriormean}{{posterior mean}}
\newcommand{\posteriorcovariance}{{posterior covariance}}

\newcommand{\prior}{{prior}}
\newcommand{\Priormean}{{Prior mean}}
\newcommand{\priormean}{{prior mean}}
\newcommand{\priormeans}{{prior means}}
\newcommand{\Priorcovariance}{{Prior covariance}}
\newcommand{\priorcovariance}{{prior covariance}}
\newcommand{\distancethreshold}{{distance threshold}}
\newcommand{\distancethresholds}{{distance thresholds}}
\newcommand{\distancecutoff}{{distance cutoff}}
\newcommand{\decaypower}{{decay power}}
\newcommand{\featurecorrelationlength}{{feature correlation length}}
\newcommand{\realspacedecaylength}{{real space decay length}}

\newcommand{\fractionalcoordinates}{{fractional coordinates}}
\newcommand{\onebodyinteraction}{{one-body interaction}}
\newcommand{\onebodyinteractions}{{one-body interactions}}
\newcommand{\onebodyinteractionuncertainty}{{one-body interaction uncertainty}}
\newcommand{\twobodyinteraction}{{two-body interaction}}
\newcommand{\twobodyinteractions}{{two-body interactions}}
\newcommand{\twobodyinteractionuncertainty}{{two-body interaction uncertainty}}
\newcommand{\oneandtwobodyinteractions}{{one-} and {two-body interactions}}
\newcommand{\oneortwobodyinteraction}{{one-} or {two-body interaction}}
\newcommand{\DFTuncertainty}{{model uncertainty}}
\newcommand{\numericalDFTuncertainty}{numerical DFT uncertainty}
\newcommand{\modelvector}{{model vector}} 
\newcommand{\modelvectors}{{model vectors}} 
\newcommand{\modelmatrixtext}{{model matrix}} 
\newcommand{\interactionvector}{{interaction vector}} 

\newcommand{\modelmatrix}{\ensuremath{\boldsymbol{X}}}
\newcommand{\posteriorcov}{\ensuremath{\boldsymbol{C_{post}}}}
\newcommand{\length}[1]{ \ensuremath{\pgfmathparse{#1/10}\pgfmathprintnumberto[precision=2]{\pgfmathresult}{\roundednumber} \roundednumber~nm} } 

\newcommand{\hilight}[1] {\color{black}{#1}\color{black}}

\newcolumntype{Y}{>{\centering\arraybackslash}X}
\newcolumntype{P}[1]{>{\centering\arraybackslash}p{#1}}

\newcommand{\myhline}{\noalign{\global\arrayrulewidth=1mm}
\arrayrulecolor{white}\hline}

\renewcommand\vec{\boldsymbol}

\journal{Computer Physics Communications}

\begin{document}

\begin{frontmatter}



\title{SAMPLE: Surface Structure Search \\ Enabled by Coarse Graining and Statistical Learning}

\author{Lukas H\"ormann, Andreas Jeindl, Alexander T. Egger, Michael Scherbela and Oliver T. Hofmann}

\address{Institute of Solid State Physics, NAWI Graz, Graz University of Technology, Petersgasse 16, 8010 Graz, Austria}

\begin{abstract}
In this publication we introduce SAMPLE, a structure search approach for commensurate organic monolayers on inorganic substrates. Such monolayers often show rich polymorphism with diverse molecular arrangements in differently shaped unit cells. Determining the different commensurate polymorphs from first principles poses a major challenge due to the large number of possible molecular arrangements. To meet this challenge, SAMPLE employs coarse-grained modeling in combination with Bayesian linear regression to efficiently map the minima of the potential energy surface. In addition, it uses ab initio thermodynamics to generate phase diagrams. Using the example of naphthalene on Cu(111), we comprehensively explain the SAMPLE approach and demonstrate its capabilities by comparing the predicted with the experimentally observed polymorphs.
\end{abstract}

\end{frontmatter}


\section{Introduction}
\label{sc:introduction}

The principal information about a given solid material is arguably the particular polymorph it forms upon crystallization. Aside from its chemical composition, the structure of a material strongly influences its thermal\cite{C2CP23439D}, mechanical\cite{wang2017relationships, upadhyay2013relationship}, optical\cite{PhysRevB.67.115212}, and electronic\cite{1367-2630-11-12-125010} properties. The ability to predict the crystal structure therefore constitutes a powerful tool to gain insight into a material's properties without producing it. Of particular interest are hereby polymorphs that form at organic/inorganic interfaces, given their prevalence in technical applications.

Current approaches to structure search in general include minima hopping\cite{goedecker2004minima}, basin hopping\cite{krautgasser2016global,wales1997global}, particle swarm optimization\cite{wang2012calypso}, (quasi-)random searches\cite{pickard2011ab,case2016convergence}, Bayesian learning\cite{todorovic2017efficient,packwood,bartok2010gaussian,li2015molecular}, genetic algorithms\cite{gator_marom,johnston2003evolving,sierka2010synergy,heiles2013global,d2008identifying}, neural networks\cite{behler2014representing}, molecular dynamics\cite{PhysRevLett.107.015701}, and stochastic optimization\cite{Schneider:xp5005, akkermans2013monte, kirkpatrick1983optimization}. However, most of these methods only work well for free molecules or bulk crystals, with a few notable exceptions for interfaces\cite{copie2014atomic, roussel2014predicting, copie2015surface, packwood2017chemical}.

In this publication, we introduce the SAMPLE approach to surface structure search. SAMPLE stands for \textit{\underline{S}urface \underline{A}dsorbate Poly\underline{m}orph \underline{P}rediction with \underline{L}ittle \underline{E}ffort} and allows to efficiently predict polymorphs at organic/inorganic interfaces from first principles. In two previous publications\cite{sample_veronika, sample_michi} we have already successfully applied parts of the SAMPLE approach. Here we present a comprehensive explanation and demonstration.

SAMPLE is a quasi-deterministic approach. This differentiates it from other structure search methods, such as basin hopping or genetic algorithms, which explore the potential energy surface stochastically. Additionally, SAMPLE provides a comprehensive overview over the relevant parts of the potential energy surface, including metastable configurations and defects. Moreover, SAMPLE can provide physical insight via the energies of the interactions between substrate and molecules as well as the molecule-molecule interactions\cite{sample_michi}.

The main challenge for any structure search method is the exponential increase of the number of polymorph candidates with the system's number of degrees of freedom. In fact, considering a continuous potential energy surface would lead to an infinite number of polymorph candidates. This circumstance is commonly referred to as configurational explosion.

To overcome said configurational explosion, SAMPLE employs an efficient and physically well-motivated coarse-grained model to discretize the potential energy surface. Within this model, a polymorph candidate corresponds to a particular arrangement of adsorbate molecules in a substrate supercell. Hereafter, we will use the term \textit{\configuration} to describe a polymorph candidate.

SAMPLE's coarse-graining reduces the number of {\configurations} to a finite, albeit large number. Evaluating their energies poses a conundrum: On the one hand, high accuracy is desired, since the energy of different polymorphs often differs by less than $20~meV$\cite{bernstein2011polymorphism, nyman2015static}. Such accuracies can only be provided by computationally expensive methods, such as dispersion-corrected DFT. On the other hand, calculating millions of energies requires  computationally cheap methods, which are often rather inaccurate.

To resolve this conundrum, SAMPLE utilizes experimental design theory and Bayesian learning to predict the coarse-grained potential energy surface. Hence, SAMPLE only requires highly accurate calculations for a training set of a few hundred {\configurations} to predict the remaining millions of {\configurations} with the same accuracy as the underlying electronic structure theory. These energies can then be combined with {\abinitiothermodynamics} to predict phase diagrams.

The paper is organized as follows: First, we discuss the key assumptions on which SAMPLE is based. Secondly, we describe the details regarding the coarse-grained modeling. Thirdly, we discuss the application of experimental design theory and Bayesian learning. Finally, we describe how we convert our results into phase diagrams using {\abinitiothermodynamics}. Throughout the text, we illustrate the  procedure on the example of naphthalene on Cu(111). Naphthalene is a small organic molecule, which forms a number of commensurate structures on the Cu(111) substrate depending on the preparation conditions\cite{forker, yamada}. Hence, it is an ideal test system to demonstrate the capabilities of SAMPLE.

\section{Basic Setup}

\hilight{SAMPLE gains its efficiency from coarse-graining the potential energy surface, which employs unit cells and {\localgeometries} as building blocks. In this section, we discuss how these building blocks arise naturally for commensurate interfaces, and how they are implemented into the SAMPLE algorithm. }

\subsection{Commensurability as Key Premise and its Physical Implications}
\label{sc:commensurablity_as_key_premise_and_its_physical_implications}

Calculating interface structures requires the use of periodic boundary conditions. These periodic boundary conditions need to account for periodicities of both, the substrate and the adsorbate. In other words, the adsorbate unit cells must be supercells of the substrate, i.e. the adsorbate layer must be \textit{commensurate} with the substrate. At organic/inorganic interfaces, commensurability occurs under specific circumstances\cite{epitaxy}, which can be exploited to limit the search space. Specifically, commensurability implies that the interactions of the adsorbate molecules with the substrate are dominant compared to the interactions between the molecules. More precisely, the potential energy surface of molecules on the surface is sufficiently corrugated, such that molecule-molecule interactions cannot significantly displace or distort molecules. This is commonly observed and well justified for small, relatively rigid molecules\cite{epitaxy}, such as aromatic molecules, which are often used in organic electronics.

\begin{figure}[H]

\setlength{\tabcolsep}{2pt}
\begin{tabular}{ m{2.0cm} m{0.4cm} m{2.0cm} m{0.45cm} m{2.0cm} }
\includegraphics[width=1.0\linewidth]{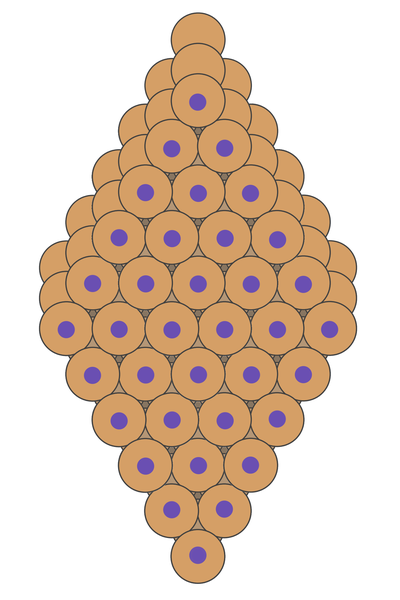} & 
\includegraphics[width=1.0\linewidth]{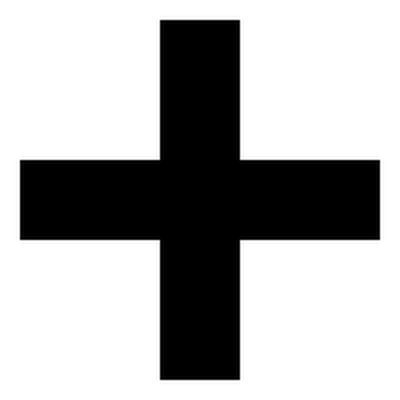} &
\includegraphics[width=1.0\linewidth]{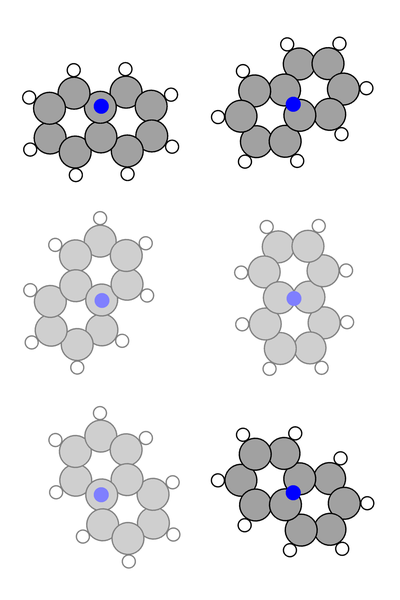} &
\includegraphics[width=1.0\linewidth]{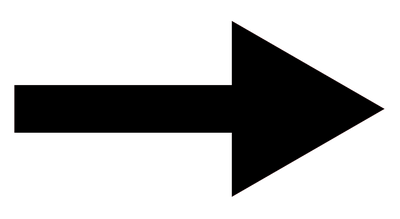} &
\includegraphics[width=1.0\linewidth]{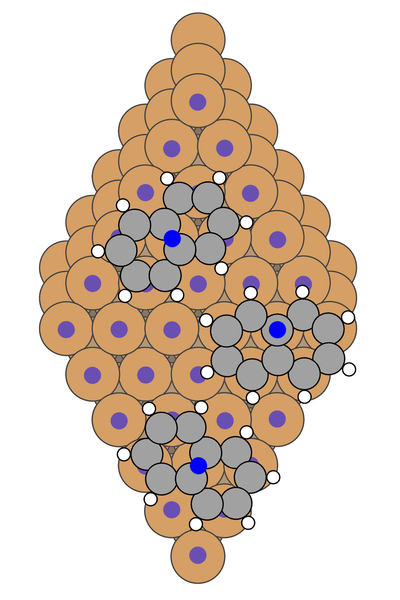}
\end{tabular}
\caption{Schematic of building a {\configuration} using a unit cell plus a number of {\localgeometries}.}
\label{fg:coarse_grained_model}
\end{figure}

The implications of commensurability enable us to develop a coarse-grained model of the potential energy surface, as illustrated in figure \ref{fg:coarse_grained_model}. First, we generate substrate supercells which serve as unit cells of the adsorbate layer. Then, we can treat the {\localgeometries} as rigid building blocks and systematically place them into unit cells to assemble an exhaustive set of {\configurations}. The geometries (see chapter \ref{sc:local_adsorption_geometries}) snap onto specific positions on the substrate (see figure \ref{fg:coarse_grained_model}). Within the coarse-grained model each {\configuration} constitutes a point on the discretized potential energy surface.

\subsection{The {\StandardUnitCell}}
\label{sc:the_standard_unit_cell}

\hilight{
	SAMPLE aims to generate all (symmetry-unique) unit cells within given constraints, such as unit cell area. However, a large set of equivalent unit cells can be generated via linear combinations of lattice vectors or symmetry transformations. {\Configurations} that form in equivalent unit cells are identical and can, therefore, be removed from the pool of structures to be calculated. Defining a convention for {\standardunitcells} allows us to identify symmetry-equivalent structures and faciliates comparision. In this section, we will describe the possible symmetry operations as well as the definition of this {\standardunitcells}.
}

\hilight{
We can represent the unit cell of any ordered adsorbate layer in fractional coordinates of the primitive substrate lattice vectors by using an epitaxy matrix $\mathbf{M}$. The epitaxy matrix allows constructing the two-dimensional super-lattice vectors $\mathbf{l_1}$ and $\mathbf{l_2}$ of the unit cell from the primitive substrate lattice vectors $\mathbf{v_1}$ and $\mathbf{v_2}$.

\begin{equation}
\begin{pmatrix} \mathbf{l_1} \\ \mathbf{l_2} \\ \end{pmatrix} = \mathbf{M} \cdot \begin{pmatrix} \mathbf{v_1} \\ \mathbf{v_2} \\ \end{pmatrix} =  \begin{pmatrix} m_1 & m_2 \\ m_3 & m_4 \\ \end{pmatrix} \cdot \begin{pmatrix} \mathbf{v_1} \\ \mathbf{v_2} \\ \end{pmatrix}
\label{eq:epitaxy}
\end{equation}

In case of a commensurate {\configuration} all elements $m_1, m_2, m_3, m_4$ of the epitaxy matrix are integer numbers.}

In principle, for every unit cell an infinite number of equivalent cells exists. By defining a set of conclusive criteria, it is possible to select one of these equivalent unit cells as the \textit{\standardunitcell}. Such a {\standardunitcell} not only allows identifying unique unit cells within the SAMPLE approach, but also to compare experimentally determined unit cells with those from SAMPLE. The criteria developed for the SAMPLE approach mainly enforce compact unit cells. We seek compactness in the oblique-angled {\fractionalcoordinates}, rather than Cartesian coordinates. This allows working directly with epitaxy matrices. Compactness is ensured by minimizing the largest diagonal of the unit cell. Additionally, the {\standardunitcell} should correspond to an epitaxy matrix that is as similar as possible to a diagonal matrix, i.e. it should have large diagonal and small off-diagonal elements. Further, we ensure that the {\standardunitcell} has right-hand chirality, i.e. the determinant of the epitaxy matrix should be positive. A detailed discussion regarding the criteria for the {\standardunitcell} can be found in chapter \ref{SI_ssc:criteria_for_the_standard_unit_cell} of the supplementary information.

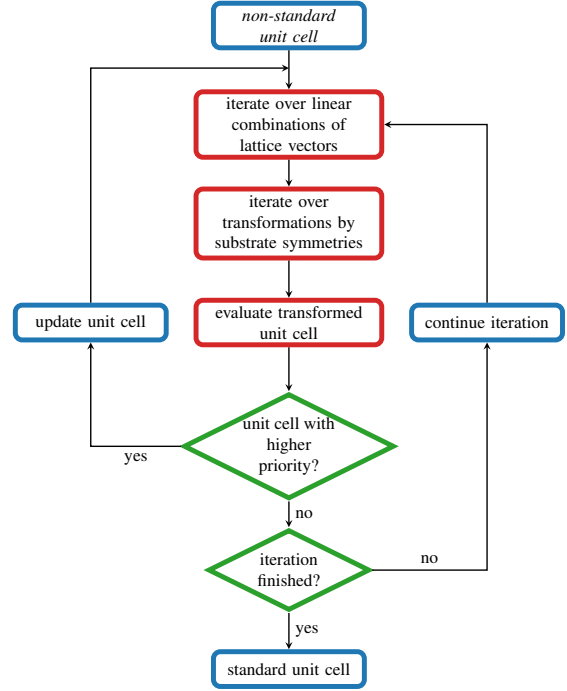
\begin{figure}[h]

\scalebox{0.7}{\begin{tikzpicture}[thick, scale=0.6, node distance = 2.0cm, auto]
    \node [block] (init) {\textit{non-standard \\ unit cell}};
    \node [iteration, below of=init] (iter3) {iterate over linear combinations of lattice vectors};
    \node [iteration, below of=iter3] (iter4) {iterate over transformations by substrate symmetries};
    \node [iteration, below of=iter4] (evaluate) {evaluate transformed unit cell};
    \node [block, left of=evaluate, node distance=4cm] (update) {update unit cell};
    
    \node [decision, below of=evaluate] (decide1) {unit cell with higher priority?};
    \node [block, right of=evaluate, node distance=4cm] (continue) {continue iteration};
    
    \node [decision, below of=decide1] (decide2) {iteration finished?};
    
    \node [block, below of=decide2] (stop) {\standardunitcell};
    \path [arrow] (init) -- (iter3);
    \path [arrow] (iter3) -- (iter4);
    \path [arrow] (iter4) -- (evaluate);
    \path [arrow] (evaluate) -- (decide1);
    \path [arrow] (decide1) -| node [near start] {yes} (update);
    \path [arrow] (update) |- ($(init)!0.43!(iter3)$) ;
    \path [arrow] (decide1) -- node {no}(decide2);
    \path [arrow] (decide2) -- node {yes}(stop);
    \path [arrow] (decide2) -| node [near start] {no}(continue);
    \path [arrow] (continue) |- (iter3);
\end{tikzpicture}}

\caption{Flow diagram of the algorithm to determine the {\standardunitcell}}
\label{fg:unit_cell_hash}
\end{figure}

\hilight{To find the {\standardunitcell} we use two types of transformations: First, combinations of lattice vectors allow generating more compact unit cells. Secondly, symmetry transformations enable orienting the unit cell such that its epitaxy matrix becomes as similar as possible to a diagonal matrix. For better handling we employ {\fractionalcoordinates} and can therefore directly combine and transform epitaxy matrices. Figure \ref{fg:unit_cell_hash} illustrates the algorithm, which we explain in more detail below.}

\hilight{
\subsubsection{Linear Combinations of Lattice Vectors}
Combining the lattice vectors $\mathbf{l_1}$ and $\mathbf{l_2}$ of an original unit cell allows generating a new unit cell with the lattice vectors $\mathbf{l_1'}$ and $\mathbf{l_2'}$.
\begin{align}
\mathbf{l_1'} ~&=~ a \mathbf{l_1} + b \mathbf{l_2}
\label{eq:combination_of_lattice_vectors_1}\\
\mathbf{l_2'} ~&=~ c \mathbf{l_1} + d \mathbf{l_2}
\label{eq:combination_of_lattice_vectors_2}
\end{align}

The new and original unit cells are equivalent if the coefficients of this combination fulfill $a, b, c, d \in \mathbb{Z}$ and if the area of the unit cell is conserved. Combinations of lattice vectors allow transforming a unit cell into a more compact shape. For mathematical convenience, we express the combination of lattice vectors (see equations \ref{eq:combination_of_lattice_vectors_1} and \ref{eq:combination_of_lattice_vectors_2}) as a transformation of the epitaxy matrix:
\begin{equation}
\begin{pmatrix} m_1' & m_2' \\ m_3' & m_4' \\ \end{pmatrix} ~=~ \underbrace{\begin{pmatrix} a & b \\ c & d \\ \end{pmatrix}}_{\vec{T}} \cdot \begin{pmatrix} m_1 & m_2 \\ m_3 & m_4 \\ \end{pmatrix}
\end{equation}

In terms of lattice vectors, the parameters $a, b, c, d$ denote how often a specific lattice vector is added. The parameters $a$ and $d$ elongate the unit cell, while $b$ and $c$ control the shear of the unit cell. To generate an equivalent unit cell, the transformation matrix $\vec{T}$ must fulfill $|\det(\vec{T})| = 1$. A detailed discussion regarding the limits for the parameters $a, b, c, d$ can be found in chapter \ref{SI_ssc:parameters_in_the_transformation_matrix_for_the_standard_unit_cell} of the supplementary information.

\subsubsection{Transformations by Substrate Point Group Symmetries}

Multiplying the lattice vectors $\mathbf{l_1}$ and $\mathbf{l_2}$ with a symmetry matrix $\vec{R}$ of the substrate transforms the unit cell into an equivalent unit cell with lattice vectors $\mathbf{l_1'}$ and $\mathbf{l_2'}$. While such transformations do not change the compactness of the unit cell, they can yield epitaxy matrices that are more similar to a diagonal matrix.
\begin{align}
\mathbf{l_1'} ~&=~ \vec{R} \cdot \mathbf{l_1} \\
\mathbf{l_2'} ~&=~ \vec{R} \cdot \mathbf{l_2}
\end{align}

\subsubsection{Chirality}
Two equivalent unit cells may also differ by the order of their epitaxy matrix elements. Switching the row in the epitaxy matrix flips the sign of the area. A unit cell with right-hand chirality has a positive area.
\begin{equation}
m_1 m_4 - m_2 m_3 > 0
\end{equation}
}

As shown in figure \ref{fg:unit_cell_hash} we find the {\standardunitcell} by iteratively applying linear combinations of lattice vectors and transformations by substrate symmetries. After each iteration step, we evaluate the priority of the transformed unit cell according to the above-specified criteria. If the new unit cell has a higher priority (better fulfills the criteria listed in chapter \ref{SI_ssc:criteria_for_the_standard_unit_cell} of the supplementary information) we restart the iteration using the current transformed unit cell as input. In case the iteration finishes without finding a better unit cell, the current unit cell is the {\standardunitcell}.

\subsection{{\LocalGeometries}}
\label{sc:local_adsorption_geometries}

The assumption of commensurability implies that inter-molecular interactions are not strong enough to expel adsorbate molecules from their local potential wells. It is therefore reasonable to expect that the geometry of an isolated molecule will not be significantly changed by the presence of other molecules, i.e. that the adsorption geometries of an isolated adsorbate molecule and a molecule in a {\closepacked} adsorbate layer are similar.

We use the term \textit{\localadsorptiongeometry} to describe the atomic structure in which an isolated molecule adsorbs on a specific position on the substrate. {\Localadsorptiongeometries} are energetically invariant under any symmetry operation of the substrate, such as rotation, mirroring or translation by primitive lattice vectors. Some symmetry operations map directly onto the original {\localadsorptiongeometry}, while other symmetry operations generate an equivalent geometry with the same adsorption energy (see figure \ref{fg:local_adsorption_geometries}). \hilight{Applying the different substrate point group symmetries to a {\localadsorptiongeometry} allows building a set of symmetry-equivalent geometries, with identical adsorption energies. Hereafter we will use the term \textit{\localgeometries} to encapsulate {\localadsorptiongeometries} and their symmetry-equivalents and use them as building blocks to generate {\configurations}. Since {\localgeometries} are invariant under translation by primitive lattice vectors, the substrate provides a natural discretization in form of the primitive substrate unit cells.}

\subsubsection{Local Geometries of Naphthalene on Cu(111)}

To find the {\localadsorptiongeometries}, we define a number of starting geometries that serve as starting points for local geometry optimizations. These optimization calculations then converge to a small number of distinct geometries, i.e. different starting geometries converge to the same final geometry. The FHI-aims quantum chemistry code (PBE + TS\textsuperscript{surf})\cite{aims, PBE, TS-surf} serves as our tool of choice to perform these calculations. Chapter \ref{SI_sc:DFT_Convergence_Test} of the supplementary information contains a description of the employed settings.

\begin{figure}[H]
\setlength{\tabcolsep}{2pt}
\begin{tabular}{m{1.25cm}|>{\centering\arraybackslash}m{1.45cm}|>{\centering\arraybackslash}m{1.45cm}|>{\centering\arraybackslash}m{1.45cm}|>{\centering\arraybackslash}m{1.45cm}}
$I, \sigma_{v,1}$ &
\includegraphics[width=\linewidth]{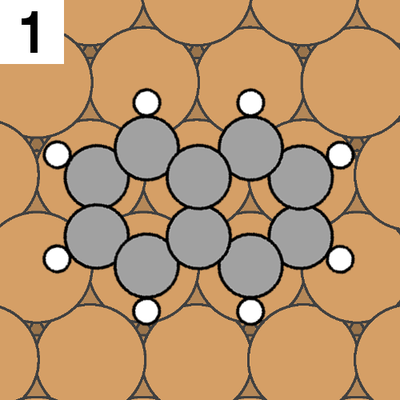} &
\includegraphics[width=\linewidth]{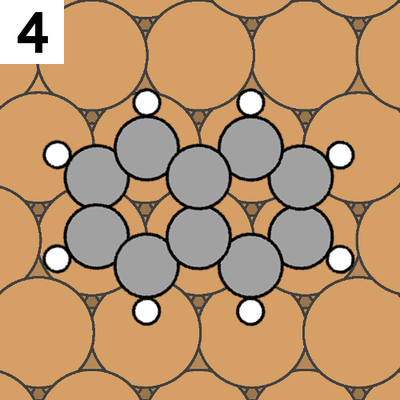} &
\includegraphics[width=\linewidth]{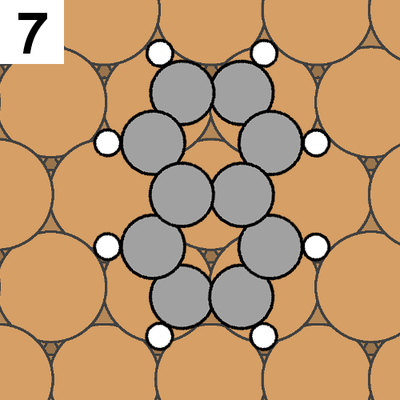} &
\includegraphics[width=\linewidth]{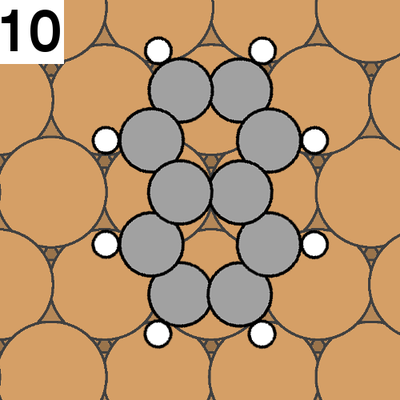} \\
$C_3, \sigma_{v,3}$ &
\includegraphics[width=\linewidth]{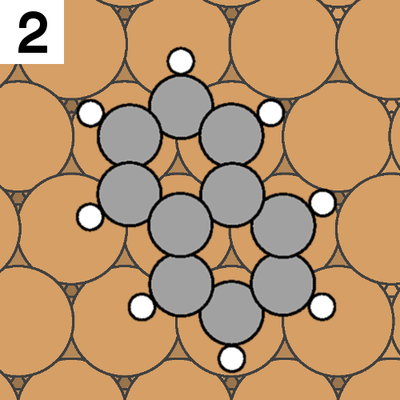} &
\includegraphics[width=\linewidth]{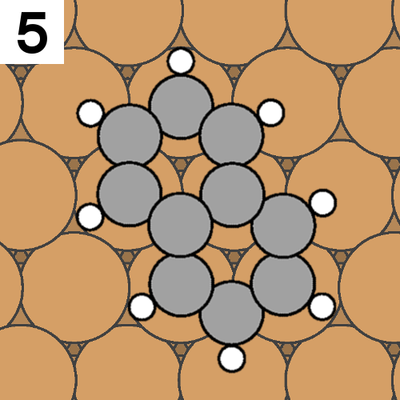} &
\includegraphics[width=\linewidth]{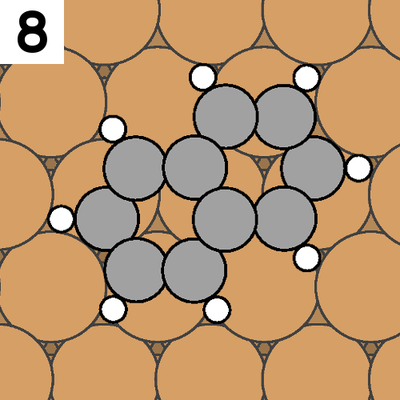} &
\includegraphics[width=\linewidth]{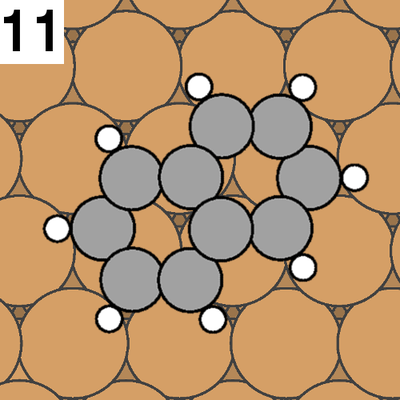} \\
$C_3^2, \sigma_{v,2}$ &
\includegraphics[width=\linewidth]{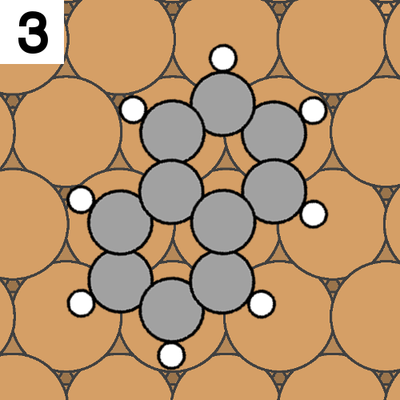} &
\includegraphics[width=\linewidth]{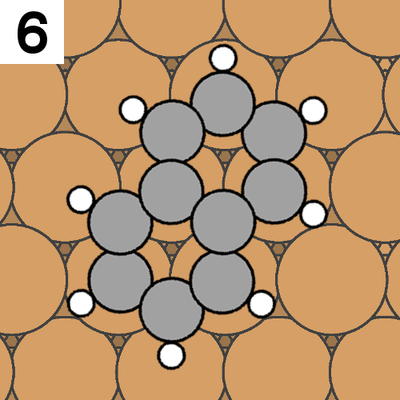} &
\includegraphics[width=\linewidth]{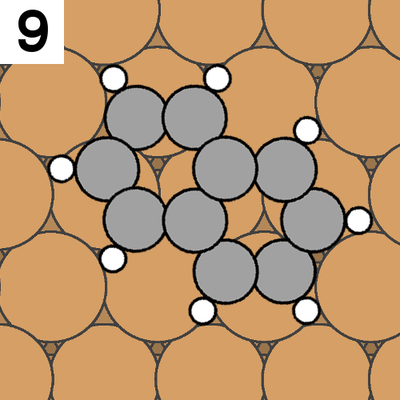} &
\includegraphics[width=\linewidth]{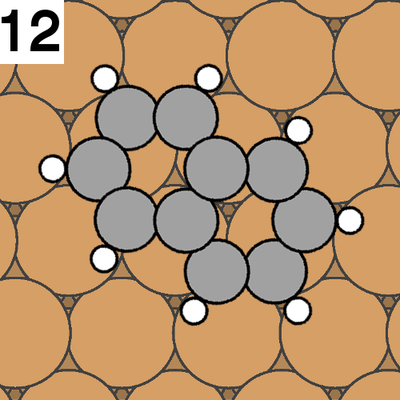} \\
&
\footnotesize{fcc hollow} &
\footnotesize{hcp hollow} &
\footnotesize{top} &
\footnotesize{bridge} \\
 &
\footnotesize{$-1.444~eV$} &
\footnotesize{$-1.421~eV$} &
\footnotesize{$-1.421~eV$} &
\footnotesize{$-1.365~eV$} \\
\end{tabular}
\caption{\hilight{{\Localgeometries} for naphthalene on Cu(111). The column shows the symmetry operation used to obtain the geometries in the respective line from the geometries in the first line.}}
\label{fg:local_adsorption_geometries}
\end{figure}

For naphthalene on Cu(111) we find four {\localadsorptiongeometries} with different adsorption energies (see figure \ref{fg:local_adsorption_geometries}). The most favorable {\localadsorptiongeometry} (fcc hollow) has an adsorption energy of $-1~444~meV$ whereas the least favorable {\localadsorptiongeometry} (bridge) has an adsorption energy of $-1~365~meV$. We find $3$ {\localgeometries} for each {\localadsorptiongeometry}, amounting to a total of $12$ {\localgeometries} of naphthalene on Cu(111), as shown in figure \ref{fg:local_adsorption_geometries}.

\section{Generating {\Configurations}}
\label{sc:generating_configuration}

The integral goal of SAMPLE is to consider all possible molecular {\configurations} in a given configuration space. The coarse-grained model SAMPLE is based upon presumes that adsorbates in {\closepacked} {\configurations} assume adsorption geometries that are similar to the {\localgeometries}. Because {\localgeometries} exist on the discrete grid of the primitive substrate lattice vectors, the number of {\configurations} in a given unit cell is finite. This circumstance enables SAMPLE to investigate all {\configurations}. \hilight{Similar to the {\standardunitcell}, we define a standard {\configuration} which we obtain by combining a {\standardunitcell} with a number of {\localgeometries}.}

\subsection{The Standard {\Configuration}}
\label{ssc:the_standard_configuration}

As previously stressed, the main challenge for structure search results from the large number of possible {\configurations}. Therefore, it would be desirable to represent {\configurations} in an efficient and easily comparable way. Since these requirements are reminiscent of the purpose hash functions have in computer science, it seems fitting to call such a representation {\configurationhash}. In simple terms, the {\configurationhash} is a one-to-one representation of a symmetry-unique {\configuration} that incorporates the epitaxy matrix, the indices of {\localgeometries} and their positions. Grouped together, these three pieces of information constitute the {\configurationhash}:
\begin{equation}
( m_1, m_2, m_3, m_4, g_1, g_2, ... , g_n, p_1, p_2, ... , p_n )
\label{eq:configuration_hash}
\end{equation}

\begin{figure}[H]
\setlength{\tabcolsep}{2pt}
\begin{tabular}{ m{2.0cm} m{3cm} m{0.45cm} m{2.0cm} }
\includegraphics[width=1.0\linewidth]{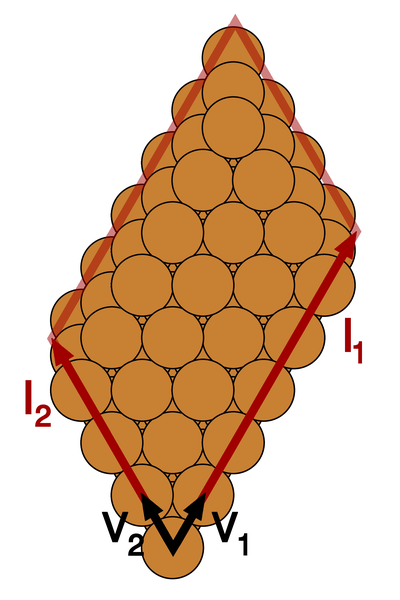} & 
\includegraphics[width=1.0\linewidth]{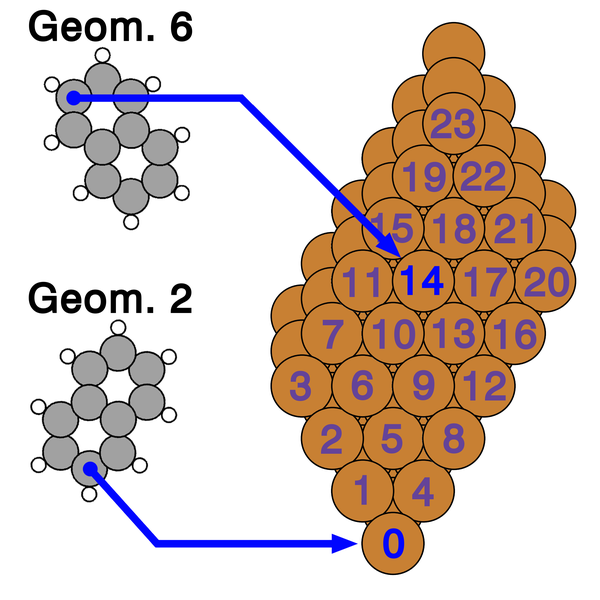} &
\includegraphics[width=1.0\linewidth]{arrow_black.png} &
\includegraphics[width=1.0\linewidth]{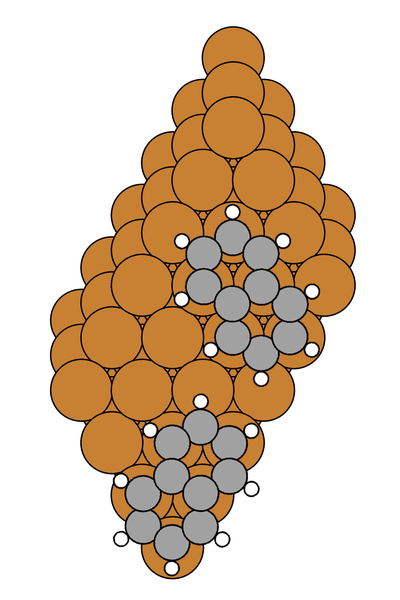} \\
$~~(6,0,0,4)$ & $~~(2,6) ~~~~~~~~~ (0,14)$ & &
\end{tabular}
\caption{\hilight{Example for the representation of a {\configuration} in form of a tuple $( 6,0,0,4,2,6,0,14 )$, whereby the elements $(6,0,0,4)$ determine the unit cell, $(2,6)$ constitute the {\localgeometries} and $(0,14)$ their positions.}}
\label{fg:configuration_hash}
\end{figure}

Here, $m_1$ thru $m_4$ are the elements of the epitaxy matrix, $g_1$ thru $g_n$ are integers that represent the {\localgeometries}. In other words, we assign each {\localgeometry} a geometry index. The final elements, $p_1$ thru $p_n$, are the position indices of the respective geometries. We construct these position indices by systematically assigning every site on the substrate an index. \hilight{Figure \ref{fg:configuration_hash} illustrates how {\configuration} and {\configurationhash} relate to each other.}

The most important function of the {\configurationhash} is to identify symmetry-unique {\configurations}. Different symmetry-equivalent {\configurations} have different tuples in the form of equation (\ref{eq:configuration_hash}). Because the {\configurationhash} must be identical for equivalent {\configurations}, we need to uniquely define one of these tuples as the {\configurationhash}. We achieve this by defining an order relation for these tuples, which is similar to alphabetic ordering of words. For example: $( 1,5 ) < ( 5,1 )$. The smallest tuple is called {\configurationhash} and the corresponding {\configuration} is the standard {\configuration}.

\subsection{Combining Local Geometries and Unit Cells}
\label{ssc:combining_equivalent_geometries_and_unit_cells}

\begin{figure}[H]
	\scalebox{0.7}{\begin{tikzpicture}[node distance = 1.9cm, auto]
		\node [block] (init1) {unit cell};
		\node [block, right of=init1,  node distance=4cm] (init2) {{\localgeometries}};
		\node [block, left of=init1, node distance=4cm] (init3) {pre-computed pair collisions};
		\node [background, below of=init1, node distance=5.8cm] (iter3) {
			\begin{tikzpicture}[level distance=1.5cm,
			level 1/.style={sibling distance=1.5cm},
			level 2/.style={sibling distance=1.5cm},
			parent anchor=south,child anchor=north,grow=south]
			\node (n0) at (-0.05,1) {build the {\configuration} tree};
			\node (n0) at (-4.2,0) [align=left, text width=3cm]{ Layer 1 };
			\node (n0) at (-4.2,-3.8) [align=left, text width=3cm]{ Layer 2 };
			\node (n0) at (-4.2,-6.5) [align=left, text width=3cm]{ $\dots$ };
			\node (n0) at (-4.2,-7.2) [align=left, text width=3cm]{ Layer $N_{ads}$ };
			\node (n1) at (-2,0) {\includegraphics[width=0.17\linewidth]{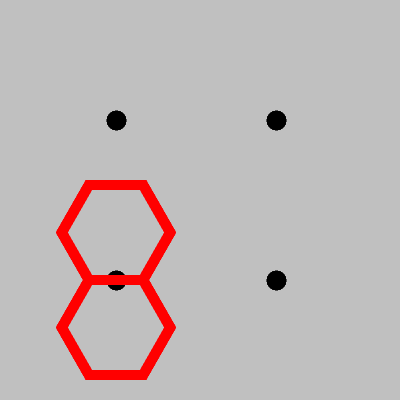}}
			child { node[anchor=north] {
					\setlength\extrarowheight{0pt}
					\begin{tabular}{c}
					\includegraphics[width=0.17\linewidth]{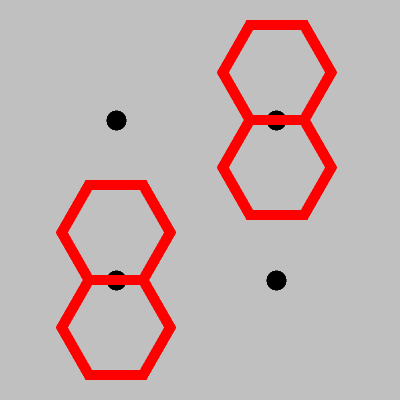} \\
					\includegraphics[width=0.17\linewidth]{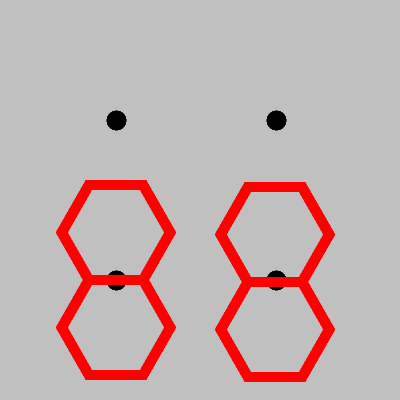} \\
					\includegraphics[width=0.17\linewidth]{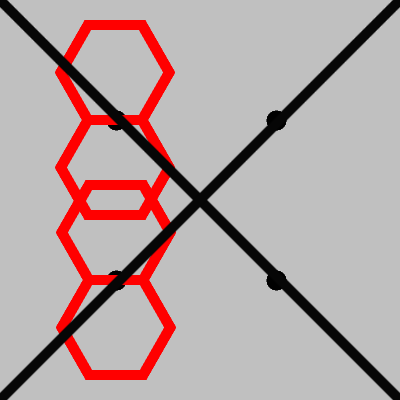}
					\end{tabular} } }
			child { node[anchor=north] {
					\setlength\extrarowheight{0pt}
					\begin{tabular}{c}
					\includegraphics[width=0.17\linewidth]{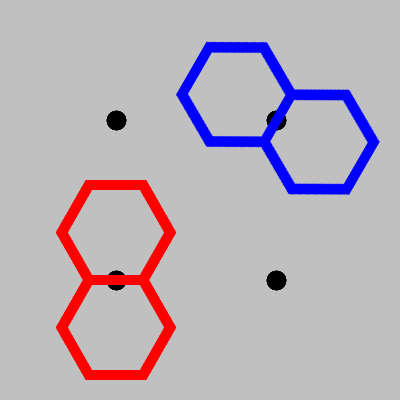} \\
					\includegraphics[width=0.17\linewidth]{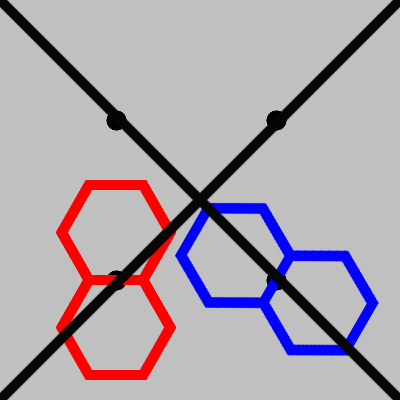} \\
					\includegraphics[width=0.17\linewidth]{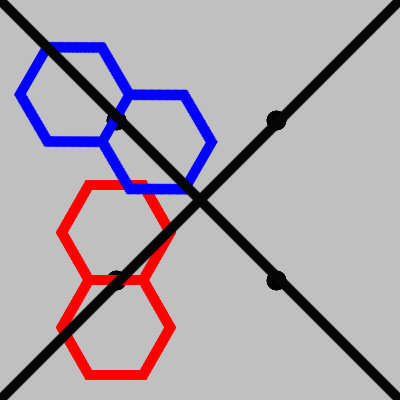}
					\end{tabular} } }
			child { node[anchor=north] {
					\setlength\extrarowheight{0pt}
					\begin{tabular}{c}
					\includegraphics[width=0.17\linewidth]{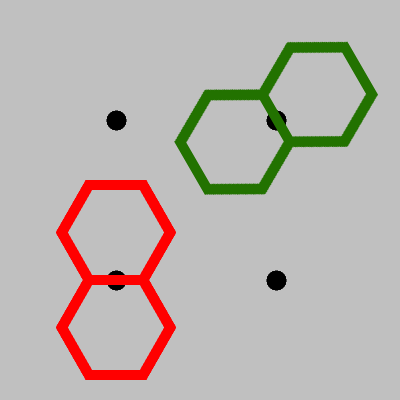} \\
					\includegraphics[width=0.17\linewidth]{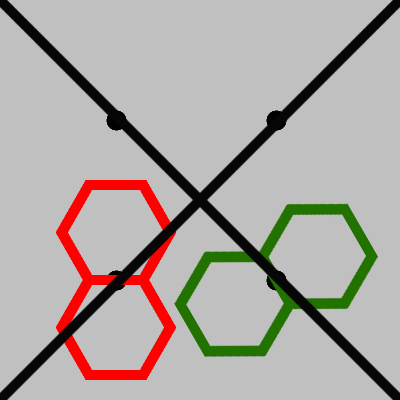} \\
					\includegraphics[width=0.17\linewidth]{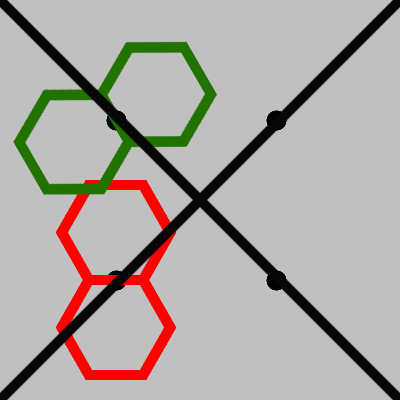}
					\end{tabular} } };
			\node (n1) at (1.9,0) {\includegraphics[width=0.17\linewidth]{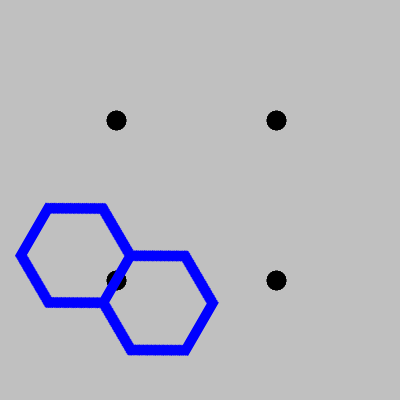}}
			child { node[anchor=north] {
					\setlength\extrarowheight{0pt}
					\begin{tabular}{c}
					\includegraphics[width=0.17\linewidth]{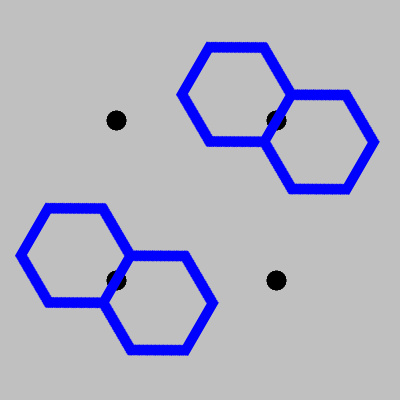} \\
					\includegraphics[width=0.17\linewidth]{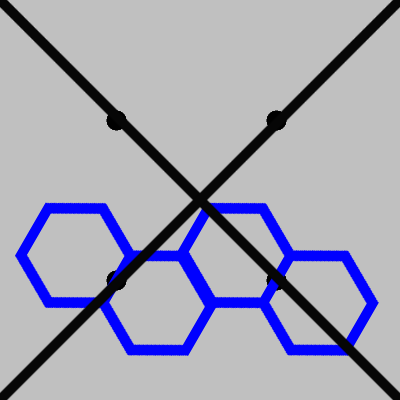} \\
					\includegraphics[width=0.17\linewidth]{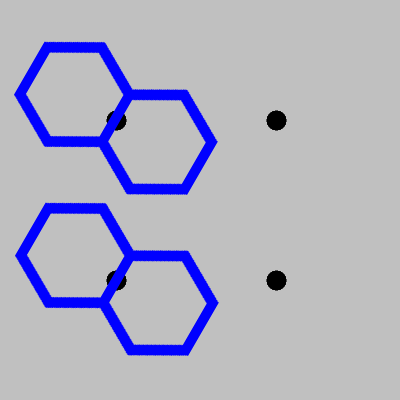}
					\end{tabular} } }
			child { node[anchor=north] {
					\setlength\extrarowheight{0pt}
					\begin{tabular}{c}
					\includegraphics[width=0.17\linewidth]{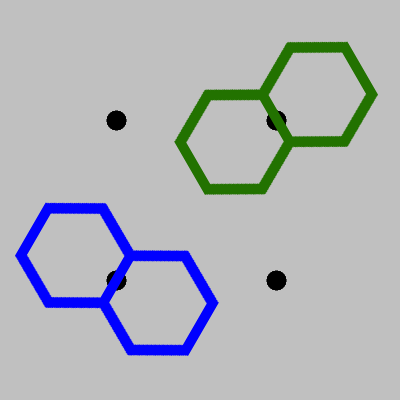} \\
					\includegraphics[width=0.17\linewidth]{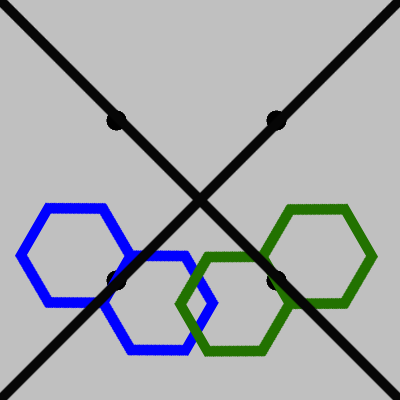} \\
					\includegraphics[width=0.17\linewidth]{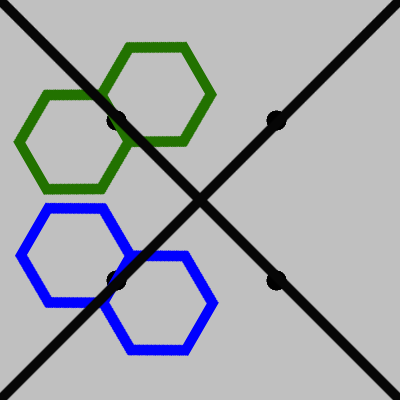}
					\end{tabular} } };
			\node (n1) at (4.3,0) {\includegraphics[width=0.17\linewidth]{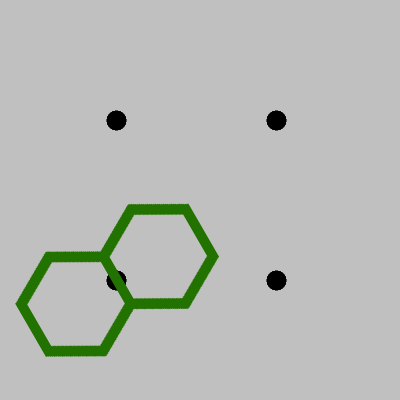}}
			child { node[anchor=north] {\setlength\extrarowheight{0pt}
					\begin{tabular}{c}
					\includegraphics[width=0.17\linewidth]{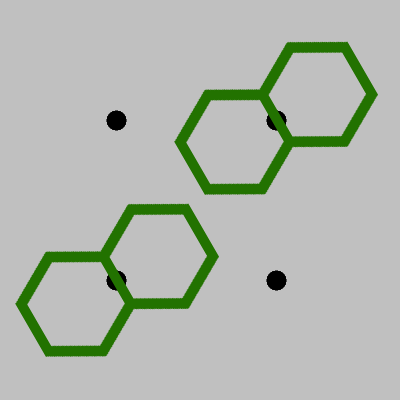} \\
					\includegraphics[width=0.17\linewidth]{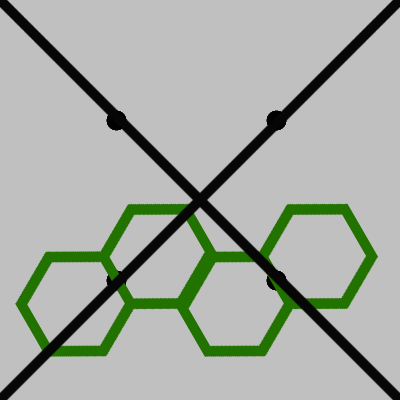} \\
					\includegraphics[width=0.17\linewidth]{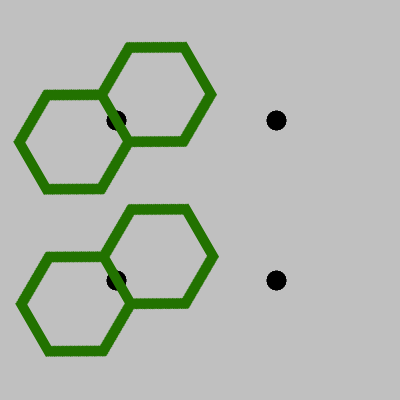}
					\end{tabular} } };
			\end{tikzpicture} };
		
		\node [redblock_wide, below of=iter3, node distance=5.8cm] (iter4) {determine {\configurationhashes} \\ remove equivalent {\configurations}};
		\node [block, below of=iter4, node distance=1.5cm] (result) {all {\configurations}};
		
		\path [arrow] (init1) -- (iter3);
		\path [arrow] (init2) -- ++(0,-0.8) -| (iter3);
		\path [arrow] (init3) -- ++(0,-0.8) -| (iter3);
		\path [arrow] (iter3) -- (iter4);
		\path [arrow] (iter4) -- (result);
		
\end{tikzpicture}}

\caption{Flow diagram of the algorithm to generate {\configurations} within one unit cell, repeating the procedure allows generating {\configurations} for multiple unit cells}
\label{fg:build_configurations}
\end{figure}

\hilight{The concept behind building a {\configuration} is simple: We place a number of {\localgeometries} onto the discretization grid that spans the unit cell and prevent collisions between adsorbates. However, comprehensive structure search requires generating all possible {\configurations}. The procedure illustrated in figure \ref{fg:build_configurations} allows generating {\configurations} in a given unit cell and, which can trivially be repeated for each unit cell from chapter \ref{sc:the_standard_unit_cell}.}

\hilight{
Generating {\configurations} requires the definition of the unit cell, the number of {\localgeometries} that should be placed in it (and thereby the coverage) as well as the set of {\localgeometries} to chose from. Additionally, we define a set of {\distancethresholds} $d_0^{AB}$, which determine the minimum distance between types of atoms belonging to different molecules. We define that two {\localgeometries} collide, if the distance between two respective atoms (of atom-type $A$ and $B$) deceeds the {\distancethreshold} $d^{AB} < d_0^{AB}$ (see figure \ref{fg:feature_vector}). This is to avoid the region of strong Pauli repulsion. Such {\configurations} are high in energy and not relevant candidates for the global minimum.

The procedure for generating {\configurations} comprises three main steps:

\begin{itemize}
	\item Pre-compute all pair collisions:\\ To improve efficiency, we predetermine a collision table containing pairs of {\localgeometries} and whether they collide. Therefore, we generate a list of all pairs of {\localgeometries} at specific distances (that fit into the unit cell). Using the {\distancethreshold} $d_0^{AB}$ (see figure \ref{fg:feature_vector}), allows determining which of these pairs collide.
	
	\item Build the {\configuration} tree:\\ The approach to systematically generate all {\configurations} takes inspiration from tree diagrams. The first layer of the tree contains all {\configurations} with one molecule per unit cell, the second layer all {\configurations} with two molecules and so on. Since {\configurations} are invariant under translations by a primitive substrate lattice vector, we place the first {\localgeometry} at the origin of the unit cell. For layer two, we add a second geometry to each {\configuration} from layer one. Hereby, we place this second geometry onto every possible position of the unit cell and check for collisions using the collision table. Adding further layers follows the same principle. The total number of layers is equal to the number of molecules per unit cell $N_{ads}$.
	
	\item Remove symmetry-equivalent {\configurations}:\\ Finally, we convert all {\configurations} into standard {\configurations} and remove any duplicates.
\end{itemize}

This procedure allows SAMPLE to generate all configurations in a given unit cell. If we wish to generate all possible configurations for a given unit cell area, or a range of areas, we repeat the configuration generation for all corresponding unit cells.
}

\subsection{{\Configurations} for Naphthalene on Cu(111)}
\label{ssc:configurations_for_naphthalene_on_cu111}

\begin{figure}[H]
\includegraphics[width=\linewidth]{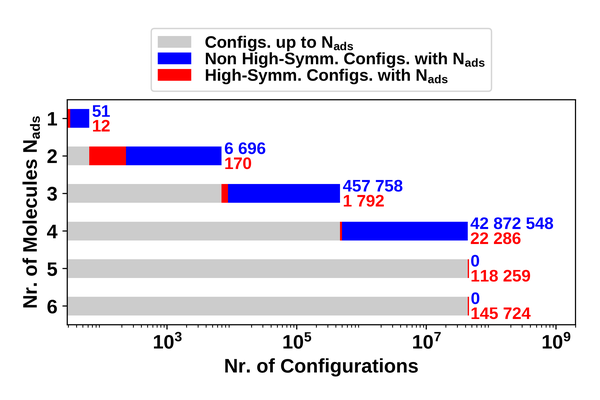}
\caption{Number of {\configurations} with increasing number of molecules per unit cell}
\label{fg:number_of_configurations}
\end{figure}

\hilight{The commensurate polymorphs of naphthalene on Cu(111), that were found in experiment, exhibit coverages of $1.19~N_{ads} nm^{-2}$ to $1.48~N_{ads} nm^{-2}$ ($15$ to $12$ primitive substrate unit cell areas $A_{PUC}$ per naphthalene molecule) and have between $1$ and $6$ flat lying molecules per unit cell\cite{forker, yamada}. For our structure prediction, we therefore select a slightly larger coverage range of $1.19$ to $1.98~N_{ads} nm^{-2}$ ($15~A_{PUC}$ to $9~A_{PUC}$ per naphthalene molecule) with $1$ to $6$ molecules per unit cell.

Since all outside atoms of naphtalene are H atoms, it is sufficient to define a single distance threshold $d^{HH}$. Here, we find that a distance of $0.15~nm$ ( between twice the covalent and the van-der-Waals radius) provides an excellent compromise that allows to generate close-packed, but avoid an excessive number of high energy {\configurations}.

While SAMPLE significantly pushes back the configurational explosion, we still eventually run into a proverbial configurational wall, as shown in figure \ref{fg:number_of_configurations}. This prohibits exhaustively generating {\configurations} with more than $4$ molecules per unit cell. Requiring {\configurations} to fulfill at least one symmetry operation of the substrate allows us to generate \textit{high-symmetry} {\configurations} with larger numbers of molecules. We generate about $43.6\cdot10^6$ {\configurations} with $1$ to $4$ molecules per unit cell and, additionally, around $25\cdot10^4$ \textit{high-symmetry} {\configurations} with $5$ and $6$ molecules per unit cell. The configurational explosion requires us to limit the coverage range for $6$ molecules to $1.37~N_{ads} nm^{-2}$ to $1.98~N_{ads} nm^{-2}$ ($13~A_{PUC}$ to $9~A_{PUC}$ per naphthalene molecule).}

\section{Energy Determination}
\label{sc:energy_determination}

For the present example, SAMPLE's coarse-graining model generates a large number (in the order of $10^7$) of possible {\configurations}. Finding thermodynamically stable {\configurations} requires ranking these possible {\configurations} with respect to their adsorption energy or Gibbs free energy. The small differences in the adsorption energy of {\configurations} necessitate high accuracy. However, directly calculating all adsorption energies is unfeasible due to the large number of possible {\configurations}.

To tackle this challenge, we employ Bayesian linear regression, which is based on a simple energy model and a number of physically motivated prior assumptions. Training the energy model with a small number of {\configurations}, whose energies are calculated with dispersion-corrected DFT, allows predicting the energies of all generated {\configurations}.

\subsection{The Energy Model}

\begin{figure}[H]
\setlength{\tabcolsep}{2pt}
\begin{tabular}{ccc}
\includegraphics[width=0.32\linewidth]{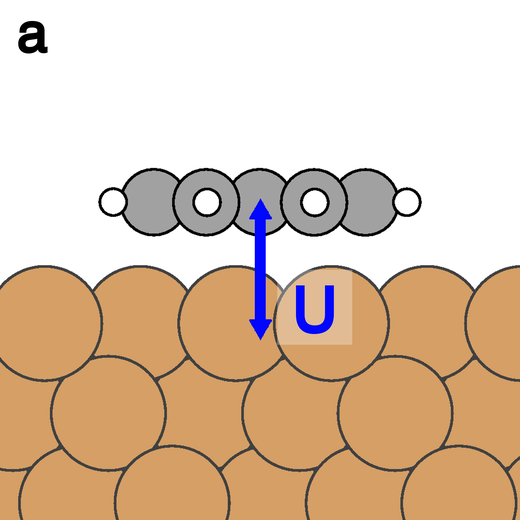} &
\includegraphics[width=0.32\linewidth]{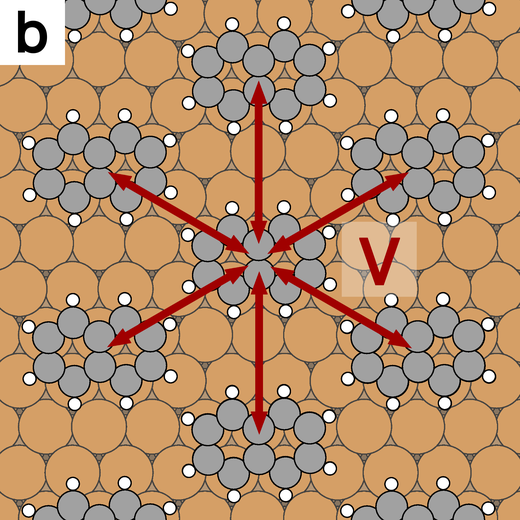} &
\includegraphics[width=0.32\linewidth]{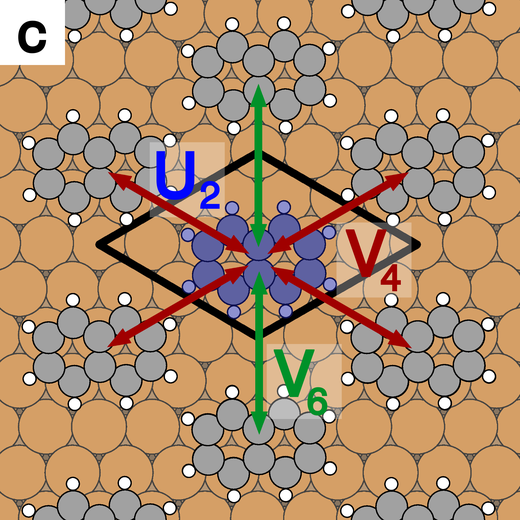}
\end{tabular}
\caption{Components of the energy model: (a) {\onebodyinteractions}, (b) {\twobodyinteractions} (c) example of the energy model for a {\configuration} with $1$ molecule per unit cell}
\label{fg:energy_model}
\end{figure}

The energy model takes its motivation from a series expansion of the adsorption energy of a {\configuration} in terms of one-, two-, and many-body interactions. For our systems it is sufficient to consider {\oneandtwobodyinteractions} (see figure \ref{fg:energy_model}). This approach is somewhat reminiscent of cluster expansion. However, cluster expansion usually also considers three-, four- and further n-body interactions. Additionally, cluster expansion methods often assume that the effective cluster interaction coefficients are sparse, i.e. that only a small fraction of clusters is relevant for describing the property of interest (e.g. energy)\cite{nguyen2017robustness}. Conversely, SAMPLE's energy model is not sparse. Considering all one and two-body contributions allows us not only to predict adsorption energies but also to extract physical insight from the energy model's coefficients. 

\hilight{
The {\onebodyinteractions} are the interactions between the adsorbate at a specific adsorption site and the substrate. We note that {\onebodyinteractions} are independent of the coverage, i.e. (de)polarization and similar effects are shifted onto the two-body terms. The {\twobodyinteractions} take account of the energy contributions of individual pairs of {\localgeometries}. The discretization employed in SAMPLE guarantees that the number of different adsorbate pairs within a given {\distancecutoff} is finite. The {\distancecutoff} is motivated by the fact that interaction energies between molecules decay to zero at large distances.
}

Using SAMPLE's energy model, we can express the energy of a {\configuration} as a sum of {\onebodyinteractions} $U_g$ and {\twobodyinteractions} $V_p$, whereby $n_g$ and $n_p$ describe how often these interactions occur.
\begin{equation}
E_{config} = \sum_{g} n_g U_g + \sum_{p} n_p V_p
\label{eq:energy_model_1}
\end{equation}

For the example in figure \ref{fg:energy_model}c $E_{config}$ contains the {\onebodyinteraction} term $U_2$ once and the {\twobodyinteraction} terms $V_4$ and $V_6$ four and two times respectively. We divide the {\twobodyinteraction} by two to account for double counting. For this example, we select a {\distancecutoff} that only includes next nearest neighbors. Note that we usually also consider more distant pairs, which we neglect for this example for the sake of clarity. Using the energy model, the adsorption energy of this example in figure \ref{fg:energy_model}c is given by:
\begin{equation}
E_{config} = 1 U_2 + 2 V_4 + 1 V_6
\label{eq:energy_model_example}
\end{equation}

Using equation (\ref{eq:energy_model_1}) allows predicting the energies of all {\configurations}, provided we know all {\oneandtwobodyinteractions}. However, it is difficult to calculate these model parameters directly. We could, for instance, use a large unit cell with periodic boundary conditions to find the {\onebodyinteraction} of an isolated {\localadsorptiongeometry}. This approach would, however, neglect collective effects, such as depolarization that results from a {\closepacked} layer. Therefore, it would be desirable to calculate energies of {\closepacked} {\configurations} and infer the {\oneandtwobodyinteractions}. This way, the aforementioned effects would be implicitly included in the energy model.

We can use equation (\ref{eq:energy_model_1}) to formulate a linear regression, which will bring us closer to the aforementioned goal. Therefore, we first rewrite equation (\ref{eq:energy_model_1}) as a vector multiplication. Note that hereafter we will consider the adsorption energy per adsorbate molecules $E = E_{config} / N_{ads}$.
\begin{equation}
E = \frac{E_{config}}{N_{ads}} = \vec{n} \cdot \vec{\omega}^T
\label{eq:energy_model_3}
\end{equation}

The {\modelvector} $\vec{n}$ lists how often each {\oneortwobodyinteraction} occurs in a given {\configuration}. Hence, elements for {\oneandtwobodyinteractions} that are not present in the {\configuration} are zero. For {\oneandtwobodyinteractions} that do occur, the respective elements in $\vec{n}$ consist of the number of occurrences divided by the number of molecules in the unit cell $N_{ads}$.
\begin{equation}
\vec{n} = \left( 0, \dots, \frac{n_g}{N_{ads}}, 0, \dots, \frac{n_p}{N_{ads}}, 0, \dots \right)
\end{equation}

For the example in figure \ref{fg:energy_model}c the {\modelvector} $\vec{n}$ would contain three non-zero entries.
\begin{equation}
\vec{n} = \left( \underbrace{0, 1, 0 \dots}_{\text{one-body}}, \underbrace{0, 0, 0, 2, 0, 1, 0, \dots}_{\text{two-body}} \right)
\end{equation}

The {\interactionvector} $\vec{\omega}$ contains the energies assigned to the {\oneandtwobodyinteractions}.
\begin{equation}
\vec{\omega} = \left( U_1, U_2, \dots, V_1, V_2, \dots \right)
\end{equation}

For a set of {\configurations}, equation (\ref{eq:energy_model_1}) becomes a matrix multiplication of a {\modelmatrixtext} $\modelmatrix$, whose lines are the vectors $\vec{n}$, and the {\interactionvector} $\vec{\omega}$.
\begin{equation}
\vec{E} = \modelmatrix \cdot \vec{\omega}^T
\label{eq:energy_model_2}
\end{equation}

\hilight{Having discussed its components, we can now formulate the aforementioned linear regression for the {\interactionvector} $\vec{\omega}$. Hereby, it is necessary to allow for some flexibility in the fitting. To so do, we assume that the model energies $\vec{E}$ can scatter around the actually calculated DFT energies $\vec{E}_{DFT}$ for a set of {\configurations} by a normal-distributed noise $\vec{\epsilon}$.
\begin{equation}
\vec{E}_{DFT} = \vec{E} + \vec{\epsilon} = \modelmatrix \cdot \vec{\omega}^T + \vec{\epsilon}   ~~~ with ~~~ \vec{\epsilon} \sim \mathcal{N}(\vec{0},\sigma_{model}^{2} \mathbbm{1} )
\label{eq:linear_regression}
\end{equation}

The noise $\vec{\epsilon}$ in equation \ref{eq:linear_regression} is determined by the {\DFTuncertainty} $\sigma_{model}$. This is the inherent uncertainty between the energy model and the DFT calculations that remains even if the model is trained on an infinite number of DFT calculations. The number of elements in the {\interactionvector} $\vec{\omega}$  (i.e, the total number of {\oneandtwobodyinteractions}) determines the dimensionality of the linear regression (see chapter \ref{sssc:dimensionality_of_the_problem}). Calculating the energies of a small number of {\configurations} $\vec{E}_{DFT}$ and trying to find the least squares solution will however fail, if the system is under-determined. This is usually the case for the SAMPLE approach, since we typically use several $100$ DFT calculations to determine several $1000$ {\oneandtwobodyinteractions}. This is solved employing the framework of Bayesian linear regression.}

\subsection{Bayesian Linear Regression}

Having formulated an energy model consisting of {\oneandtwobodyinteractions}, we now aim to determine these interactions with the aid of Bayesian linear regression, which is given by equation \ref{eq:bayes}:

\begin{equation}
p(\vec{\omega}|\vec{E}_{DFT}) = \frac{ p(\vec{E}_{DFT}|\vec{\omega}) ~ p(\vec{\omega}) }{p(\vec{E}_{DFT})}
\label{eq:bayes}
\end{equation}

Bayes' theorem enables expressing the unknown {\posterior} for the {\oneandtwobodyinteractions} $p(\vec{\omega}|\vec{E}_{DFT})$ in terms of two known probability distributions. These are the {\likelihood} $p(\vec{E}_{DFT}|\vec{\omega})$ and the {\prior} $p(\vec{\omega})$. In addition, the \textit{marginal probability} $p(\vec{E}_{DFT})$ normalizes the {\posterior}. Here the \textit{marginal probability} is the probability to find DFT energies $\vec{E}_{DFT}$.

The {\likelihood} is the probability to get $\vec{E}_{DFT}$ given certain {\oneandtwobodyinteractions} $\vec{\omega}$. It directly follows from equation (\ref{eq:linear_regression}), whereby $\sigma_{model}$ is the {\DFTuncertainty}, i.e. the inherent uncertainty between the energy model and DFT.
\begin{equation}
p(\vec{E}_{DFT}|\vec{\omega}) \propto \exp \left( -{\frac{1}{2 \sigma_{model}^2} || \vec{E}_{DFT} - \modelmatrix\vec{\omega} ||^2 } \right)
\end{equation}

The {\prior} allows including physical knowledge about the system and thereby enables finding the expectation value for the {\oneandtwobodyinteractions}, even if the problem posed by equation \ref{eq:linear_regression} is under-determined. The {\priorcovariance} matrix $C$ couples the parameters in $\vec{\omega}$ and thereby reduces the number of independent parameters. We can write the {\prior} as a normal distribution with the {\priormean} $\vec{\omega}_0$ and the {\priorcovariance} matrix $C$:
\begin{equation}
p(\vec{\omega}) \propto \exp \left( -{ \frac{1}{2} (\vec{\omega} - \vec{\omega}_0)^T \vec{C}^{-1} (\vec{\omega} - \vec{\omega}_0) } \right)
\end{equation}

Since both the {\likelihood} and the {\prior} are Gaussians, {\posterior} is also a Gaussian. Further, the marginal probability $p(\vec{E}_{DFT})$ is taken to be constant and can therefore be neglected. Hence, we write the {\posterior} as follows:
\begin{equation}
p(\vec{\omega}|\vec{E}_{DFT}) \propto \exp \left( -{ \frac{1}{2} (\vec{\omega} - \bar{\vec{\omega}})^T \posteriorcov^{-1} (\vec{\omega} - \bar{\vec{\omega}}) } \right)
\label{eq:posterior}
\end{equation}

Here, the new {\posteriorcovariance} matrix has the following expression:
\begin{equation}
\posteriorcov^{-1} = \frac{ \modelmatrix^T \modelmatrix }{ \sigma_{model}^2} + \vec{C}^{-1}
\label{eq:covariance_martix_A}
\end{equation}

The {\posteriormean} $\bar{\vec{\omega}}$ is also the expectation value for the {\interactionvector} $\omega$, which contains the {\oneandtwobodyinteractions}.
\begin{equation}
\bar{\vec{\omega}} = \posteriorcov \left( \frac{ \modelmatrix^T E_{DFT} }{ \sigma_{model}^2} + \vec{C}^{-1} \vec{\omega}_0 \right)
\label{eq:posterior_mean}
\end{equation}

Equation (\ref{eq:posterior_mean}) is the best estimator for the {\interactionvector} $\omega$. To fit the energy model, we supply a set of {\configurations} with energies from DFT and calculate the {\oneandtwobodyinteractions}. This in turn allows predicting the energies of all {\configurations}.

\subsubsection{\Priormean}

The {\priormean} $\vec{\omega}_0$ takes account of the prior knowledge regarding the {\oneandtwobodyinteractions}. For the {\onebodyinteractions}, the adsorption energy of a {\localadsorptiongeometry}, i.e. an isolated molecule, provides a convenient estimator. Additionally, the {\onebodyinteractionuncertainty} $\sigma_{one-body}$ controls how much a {\onebodyinteraction} may differ from its {\prior}.

Determining a {\priormean} for the {\twobodyinteractions} is less straightforward. Since their energy contribution can either be attractive or repulsive, an obvious choice is starting with non-interacting geometries with a {\priormean} of zero.

In certain cases it is useful to employ SAMPLE in a two step approach and start with a two-body interaction prior from the interactions in the gas phase, rather than assuming non-interaction molecules on the surface. In this case, we first determine {\twobodyinteractions} for the adsorbate layer in vacuum (using a non-interacting {\priormean}). Since the substrate requires the bulk of the computational effort of a DFT calculation, removing it allows generating a larger amount of data and consequently a good fit for the {\twobodyinteractions}. We then use these {\twobodyinteractions} for a molecule sheet in vacuum as {\priormean} for the adsorbed system. Such an approach works well for molecules with comparably strong molecule-molecule interactions, such as benzoquinone, which forms hydrogen bonds.

\begin{figure}[H]
\includegraphics[width=1\linewidth]{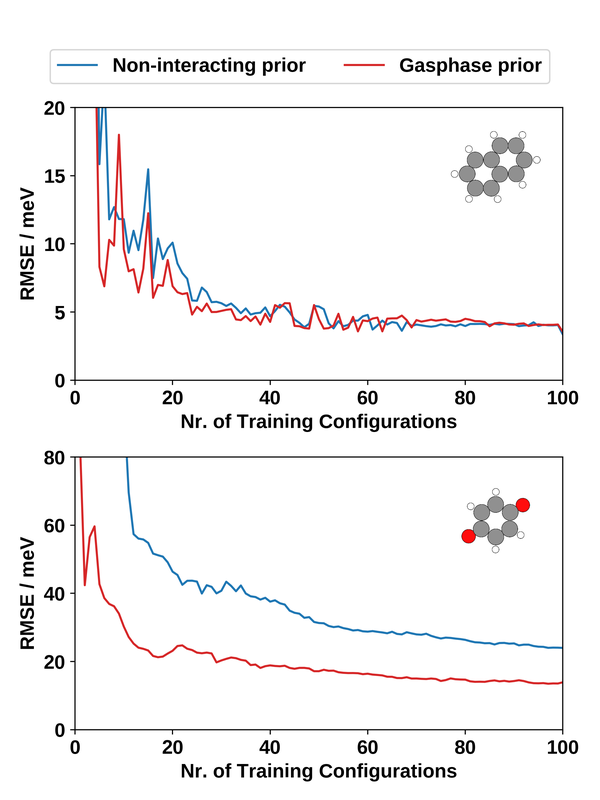}
\caption{Dependence of the root mean square error of the predicted adsorption energies per molecule on the number of training {\configurations} (see chapter \ref{ssc:D-optimal_selection}) for non-interacting {\priormean} and gas phase {\priormean}. Top: Naphthalene on Cu(111). Bottom: Benzoquinone on Ag(111)}
\label{fg:gasphase_vs_zero_prior}.
\end{figure}

However, molecules on the substrate can experience charge transfer, leading to electrostatic repulsion and therefore significantly different interactions compared to a molecule sheet in vacuum. This is indeed the case for Naphthalene on Cu(111), where the work function of the adsorbate system is about $1~eV$ lower than that of the clean Cu(111) slab. This work function shift indicates that the Naphthalene molecules become positively charged upon adsorption on the Cu substrate. Figure \ref{fg:gasphase_vs_zero_prior} shows a comparison of the performance of both {\priormeans} for naphthalene on Cu(111) and benzoquinone on Ag(111). For relevant training set sizes ($N_{set} > 60$) a gas phase {\priormean} does not improve the prediction for naphthalene, while we find significantly better predictions for benzoquinone. Increasing the number of training point leads to a convergence of the predictions with non-interacting {\priormean} and gas phase {\priormean}.

\subsubsection{{\Priorcovariance}}

The {\priorcovariance} enables including correlations between different interactions. The SAMPLE approach assumes no correlation between different {\onebodyinteractions} or between {\oneandtwobodyinteractions}, but non-zero correlations between different {\twobodyinteractions}. The {\priorcovariance} for {\twobodyinteractions} rests upon the assumption that similar pairs of geometries have similar {\twobodyinteractions}. Note that similarity is not physically defined per se, but must be encoded via a (more or less) arbitrary {\featurevector} (see chapter \ref{ssc:the_feature_vector}). Furthermore, the {\priorcovariance} is constructed in such a way that it converges to zero at large distances. In combination with the {\priormean}, this accounts for the assumption that the {\twobodyinteractions} decays to zero at large distances. We use a multiplicative exponential kernel, which consists of two contributions:

\begin{itemize}
\item First, the distance in feature space $|\vec{f}_i - \vec{f}_j|$ provides a measure for the similarity between pairs, whereby $\vec{f_i}$ is the {\featurevector} belonging to a specific pair of molecules. We also introduce the {\featurecorrelationlength}, which is used as a reference length to scale distances $|\vec{f}_i - \vec{f}_j|$ in the feature space. A large {\featurecorrelationlength} $\xi$ entails non-disappearing correlation even for less similar pairs, while a small $\xi$ restricts correlation to very similar pairs. In other words, if the feature distance $|\vec{f}_i - \vec{f}_j|$ is small compared to $\xi$, we assume similar {\twobodyinteractions} for geometry pair $i$ and geometry pair $j$. In case the feature distance is large compared to $\xi$, the {\twobodyinteractions} may vary widely. Note that for {\onebodyinteractions} only the $C_{ii}$ are non-zero.
\begin{equation}
C_{ij} = \sigma_i^* \sigma_j^* \exp \left( - \frac{|\vec{f}_i - \vec{f}_j|}{\xi} \right)
\label{eq:BLR_cov}
\end{equation}

\item Second, we introduce a decay of the {\twobodyinteractions} at large distances. This term depends on the minimum distance $d_{min,i}^{AB}$ between the atoms A-B of two geometries. Additionally, it includes the {\distancethreshold} $d_0^{AB}$ and the {\realspacedecaylength} $\tau^{AB}$ for the pair atom species A-B. To attain a single value $\sigma_i^*$ we take the arithmetic mean of the different atom-species pairs $A$-$B$. Note that other means may constitute an equally valid choice.
\begin{equation}
\sigma_i^* = \sigma_{two-body} \frac{1}{N_{AB}} \sum_{AB} \exp \left(- \frac{d_{min,i}^{AB} - d_0^{AB}}{\tau^{AB}} \right)
\label{eq:BLR_std}
\end{equation}

$\sigma_{two-body}$ is the standard deviation of the {\prior}. The parameter $\tau^{AB}$ controls at which distance the {\twobodyinteractions} converge to their {\priormean} $\omega_0$. A small $\tau^{AB}$ leads to a steeper decay to the {\priormean}, while a large $\tau^{AB}$ leads to a slower decay. Given a fixed $\tau^{AB}$, small distances between geometries ($|d_{min,i}^{AB} - d_0^{AB}| \ll \tau^{AB}$ with $d_{min,i}^{AB} - d_0^{AB} \geq 0$) result in a large $\sigma_i^*$, allowing for a larger difference between the {\twobodyinteractions} and their {\priormean}. For large distances $\sigma_i^*$ becomes small, allowing only for a small difference between interactions and {\priormean}.

\end{itemize}

\hilight{
\subsubsection{Dimensionality of the Problem}
\label{sssc:dimensionality_of_the_problem}

The dimensionality of the problem at hand depends on the size of the {\interactionvector} $\vec{\omega}$, which is given by the number of {\oneandtwobodyinteractions}. The number of {\onebodyinteractions} 
is equal to the number of {\localadsorptiongeometries} (four in case of naphthalene on Cu(111)). The number of {\twobodyinteractions} depends, roughly speaking, on the number of (non-colliding) pairs. More precisely, it is determined by the {\distancecutoff} and the settings for the {\featurevector} (see chapter \ref{ssc:the_feature_vector}).

The {\priorcovariance} (see equation \ref{eq:BLR_cov}) defines a correlation between the {\twobodyinteractions}. Hereby we may loosely differentiate between weakly and strongly correlated {\twobodyinteractions}. While weakly correlated {\twobodyinteractions} constitute essentially independent degrees of freedom, highly correlated {\twobodyinteractions} do not increase the number of degrees of freedom.

In SAMPLE, the number of weakly correlated {\twobodyinteractions} converges to a finite number. While changing the settings for the {\featurevector} and increasing the {\distancecutoff} allows to significantly increase the total number of {\twobodyinteractions}, many of them are highly correlated.

Using more {\twobodyinteractions} than necessary, while keeping otherwise similar settings, neither improves nor worsens the prediction accuracy (see chapter \ref{SI_ssc:dimensionality_of_the_interactionVector} in the supplementary information). Typical systems require several $1000$ {\oneandtwobodyinteractions}.
}

In SAMPLE, the number of weakly correlated \twobodyinteractions converges to a finite number. While the total number of \twobodyinteractions depends on the precise settings of the feature vector (i.e., its dimensionality, the feature threshold, and the distance cutoff), increasing these values eventually only generate strongly correlated fit parameters. 

While a large number can 
This can be seen by hypothetically increasing the distance cutoff $\tau$. As can be seen from eq. WHATEVER, this may add additional \twobodyinteractions, but these interaction are strongly correlated with interactions SOMETHING. 

Since the number of weakly correlated {\twobodyinteractions} is limited, increasing the dimensionality of the {\interactionvector} will eventually only increase the number of highly correlated {\twobodyinteractions}. Using more {\twobodyinteractions} than necessary, while keeping otherwise similar settings, neither improves nor worsens the prediction accuracy (see chapter \ref{SI_ssc:dimensionality_of_the_interactionVector} in the supplementary information). Typical systems requires several $1000$ {\oneandtwobodyinteractions}.

\subsubsection{Hyperparameters}

Bayesian linear regression requires a number of hyperparameters, namely the {\onebodyinteractionuncertainty} $\sigma_{one-body}$, the {\twobodyinteractionuncertainty} $\sigma_{two-body}$, the {\DFTuncertainty} $\sigma_{model}$, the {\featurecorrelationlength} $\xi$, the {\realspacedecaylength} $\tau^{AB}$ and the {\distancecutoff} $d^{AB}_{max}$.

The {\twobodyinteractionuncertainty} $\sigma_{two-body}$ is the square root of the variance between {\priormean} and learned interaction energy. This hyperparameter should have a value in the range of the {\twobodyinteractions} at molecule distances close to the {\distancethreshold}. For the {\DFTuncertainty} $\sigma_{model}$ we use the {\numericalDFTuncertainty}, which we derived from the convergence criteria of our DFT settings. This is justified, since $\sigma_{model}$, which describes the scattering of the DFT energies around the energy model, is comparable to the numerical convergence of the DFT calculations.

\hilight{We use a physically motivated settings for the {\realspacedecaylength} $\tau^{AB}$ and the {\distancecutoff} $d^{AB}_{max}$. Additionally, chapter \ref{SI_sc:BLR_Convergence_Test} of the supplementary information contains convergence plots for both parameters.} \hilight{We find that the root mean square error converges for $\tau \leq 0.5~nm$ and for $d_{max} \leq 1~nm$.}  \hilight{While the {\distancecutoff} does not enter into the equations, it controls which {\twobodyinteractions} the model considers. The energy model only includes {\twobodyinteractions} of molecule pairs if the smallest distance between their respective atoms is smaller than the {\distancecutoff} $d^{AB}_{max}$ ($A$ and $B$ being the atom species).}

For the feature decay length $\xi$ it is difficult to find physically motivated values. We determine $\xi$ by minimizing the root mean square error for the naphthalene test system described in chapter \ref{ssc:D-optimal_selection} (details in chapter \ref{SI_sc:BLR_Convergence_Test} in the supplementary information). The final settings we use for naphthalene on Cu(111) are summarized in table \ref{tb:BLR_params}.

\begin{table}[H]
\small
\caption{Hyperparameter settings for Bayesian linear regression}
\label{tb:BLR_params}
\begin{tabularx}{\linewidth}{p{4.2cm}YY}
\rowcolor{lightgray}
\textbf{parameter} & \textbf{symbol} & \textbf{setting} \\
\noalign{\global\arrayrulewidth=1mm}
\arrayrulecolor{white}\hline
\rowcolor{lightgray}
{\onebodyinteractionuncertainty} & $\sigma_{one-body}$ & $100 ~meV$ \\
\rowcolor{lightgray}
{\twobodyinteractionuncertainty} & $\sigma_{two-body}$ & $100 ~meV$ \\
\rowcolor{lightgray}
{\DFTuncertainty} & $\sigma_{model}$ & $5 ~meV$ \\
\rowcolor{lightgray}
{\featurecorrelationlength} & $\xi$ & $10$ \\
\rowcolor{lightgray}
{\realspacedecaylength} & $\tau^{HH}$ & $\length{5}$\\
\rowcolor{lightgray}
{\distancecutoff} & $d^{HH}_{max}$ & $\length{16.0}$ \\
\end{tabularx}
\end{table}

\subsection{The {\FeatureVector}}
\label{ssc:the_feature_vector}

An integral part in SAMPLE's implementation of Bayesian linear regression is the {\featurevector}. The {\featurevector} acts as a descriptor for a pair of {\localgeometries}. Its main purpose is defining a measure of similarity between different geometry pairs. In general, the {\featurevector} should be derived from the structure (not its properties) and it must be related to the property of interest. In other words, similar {\featurevectors} should entail similar energies.

\begin{figure}[H]
	\includegraphics[width=1\linewidth]{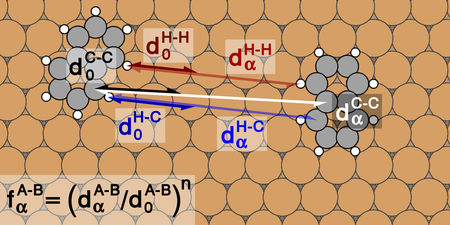}
	\caption{\hilight{Example of the three possible types of distances and the respective {\distancethresholds} for naphthalene molecules: between H atoms $d_{\alpha}^{HH}$ and $d_0^{HH}$, between H atoms and C atoms $d_{\alpha}^{CH}$ and $d_0^{CH}$, between C atoms $d_{\alpha}^{CC}$ and $d_0^{CC}$.}}
	\label{fg:feature_vector}
\end{figure}

We construct the {\featurevector} from the distances between the respective atoms of two {\localgeometries}. We differentiate between different atom species, since for instance, distances between H atoms relate differently to the adsorption energy than distances between C atoms. For naphthalene we can determine distances between H atoms, distances between H atoms and C atoms, as well as distances between C atoms. These different distances are shown in figure \ref{fg:feature_vector}.

The {\featurevector} itself consists of groups belonging to specific pairs A-B of atom species. A general {\featurevector} for naphthalene would contain groups for H-H, H-C and C-C atom pairs. Note that for the specific example of naphthalene, distances between H atoms are sufficient.
\begin{equation}
\vec{f} = \left( {\color{C3} f_1^{HH}, f_2^{HH}, \dots}, {\color{C0} f_1^{HC}, f_2^{HC}, \dots}, {\color{C7} f_1^{CC}, f_2^{CC}, \dots} \right)
\end{equation}

Each of these groups contains a number of {\featurevector} elements$f_{\alpha}$, which we calculate as follows:
\begin{equation}
f_{\alpha}^{AB} ~=~ \left( \frac{d_{\alpha}^{AB}}{d_{0}^{AB}} \right)^n
\end{equation}

Here $d_{0}^{AB}$ is the same {\distancethreshold} we used in chapter \ref{ssc:combining_equivalent_geometries_and_unit_cells}, $d_{\alpha}^{AB}$ is the distance between two atoms, and $n$ is a negative {\decaypower}. The design of the {\featurevector} is physically motivated. Since {\twobodyinteractions} decay with distance (accounted for by the {\realspacedecaylength}, see equation \ref{eq:BLR_std}), the energy contributions of close pairs vary more strongly with geometric differences than the small energy contributions of distant pairs. We account for this behavior by the negative {\decaypower} $n$. Hence, close pairs of molecules lead to large values in the {\featurevector}, whereas distant pairs lead to small values. Therefore, small geometric differences of close molecule pairs entail large differences of the {\featurevectors} and consequently less correlation. At the same time, features of distant molecule pairs show small variations.

Within each group $AB$ the elements $f_{\alpha}^{AB}$ are arranged in descending order of their magnitude. Hence, the first element of each section represents the smallest distance between the atoms $AB$ of this section. Since the {\twobodyinteractions} decay with distance, the first elements of the {\featurevector} are most relevant for representing an interacting pair of molecules. For small molecules it is possible to use all distances. For larger molecules, we can reduce the dimension of the {\featurevector} by truncating unimportant elements. To this aim, we define a {\featuredimension} determining the number of elements in each section of the {\featurevector}. \hilight{Note that the dimensionality of the {\interactionvector} only indirectly depends on the {\featuredimension} and may not even change when increasing the {\featuredimension}. The  {\featuredimension} should be as large as necessary for distinguishing between different pairs of molecules and as small as possible to limit computational effort. In case of two naphthalene molecules, there are $64$ possible distances for H atoms alone. We show in chapter \ref{SI_ssc:feature_dimension} of the supplementary information that a {\featuredimension} of $16$ already converges the root mean square error (RMSE).}

Finally, we define a {\featurethreshold} $\Delta f$, which determines whether two {\featurevectors} are similar. We consider two {\featurevectors} $\vec{f}$ and $\vec{g}$ identical, if all individual elements differ by less than a {\featurethreshold} $| f_{\alpha} - g_{\alpha} | < \Delta f$. We note that different {\localgeometry} pairs can have the same {\featurevector}.

\subsubsection{The {\featurevector} for Naphthalene}

Since the outermost atoms of a flat-lying naphthalene molecule are the H atoms, it is sufficient to consider only distances between H atoms in the {\featurevector}. Hence, we only require one {\distancethreshold} $d_{0}^{HH}$, which is the same as in chapter \ref{ssc:configurations_for_naphthalene_on_cu111}. In a pair of naphthalene molecules only $4$ H atoms of one molecule can come close to $4$ H atoms of the other molecule. Therefore, we use a {\featuredimension} of $16$, which accounts for all pair-wise distances between these $4 \cdot 4$ H atoms. Further, the {\decaypower} of $n=-2$ is inspired by the decay of Coulomb interactions. Finally, we find that a {\featurethreshold} $\Delta f=0.01$ converges the root mean square error for the naphthalene test system described in chapter \ref{ssc:D-optimal_selection} (details in chapter \ref{SI_sc:BLR_Convergence_Test} of the supplementary information). We demonstrate in chapter \ref{ssc:testing_BLR} that these setting yield excellent results. An overview of the parameter settings for the naphthalene {\featurevector} is given in table \ref{tb:feature_vector_params}.

\begin{table}[h]
\small
\caption{Hyperparameter settings for the naphthalene {\featurevector}}
\label{tb:feature_vector_params}
\begin{tabularx}{\linewidth}{p{3cm}YY}
\rowcolor{lightgray}
\textbf{parameter} & \textbf{symbol} & \textbf{setting} \\
\noalign{\global\arrayrulewidth=1mm}
\arrayrulecolor{white}\hline
\rowcolor{lightgray}
{\distancethreshold} & $d_{0}$ & $\length{1.5}$ \\
\rowcolor{lightgray}
{\featurethreshold} & $\Delta f$ & $0.01$ \\
\rowcolor{lightgray}
{\featuredimension} & & $16$ \\
\rowcolor{lightgray}
{\decaypower} & $n$ & $-2$\\{\color{white}\vrule}
\end{tabularx}
\end{table}

\subsection{Training Set Selection}
\label{ssc:D-optimal_selection}

\hilight{In order to determine the {\oneandtwobodyinteractions}, the Bayesian learning algorithm requires a set of training data. Naively, we could randomly select a number of {\configurations} from all possible {\configurations}. However, different training sets contain different amounts of information. For instance, a training set containing only {\configurations} where the molecules are very close would yield accurate predictions for such closely packed {\configurations}, but contain little information regarding loosely packed {\configurations}. Hence, predictions for such {\configurations} would be unreliable, making this training set unsuited for such a task.

A training set (see chapter \ref{ssc:D-optimal_selection}) should contain as many different features as possible. Additionally, the unit cells in the training set should be as small as possible to minimize the computational effort. Conveniently, the two-molecule {\configurations} already contain all combinations of two {\localgeometries} in all possible relative positions that fit into the unit cells. Hence, if $d_{max}$ is sufficiently small, or the unit cells with two molecules are sufficiently large, the two-molecule {\configurations} contain all possible molecule pairs and therefore all possible features.
}

The information contained in a training set is reflected by the {\posteriorcovariance} matrix $\posteriorcov$. In equation (\ref{eq:posterior}), the covariance matrix determines the uncertainty of the mean vector $\bar{\vec{\omega}}$. Hence, minimizing $\posteriorcov$ will improve the accuracy of the fit coefficients. Since an order relation for matrices is not defined, minimizing $\posteriorcov$ cannot be done directly. Experimental design theory provides a number of optimality criteria for gauging the information contained in a matrix $\posteriorcov$. The criterion chosen for SAMPLE is D-optimality\cite{wald1943}. D-optimality seeks to minimize the determinant of $\posteriorcov$. More precisely, it minimizes the generalized variance of the fit parameters.
\begin{equation}
\min \{ \det{(\posteriorcov)} \}
\end{equation}

Equation (\ref{eq:covariance_martix_A}) shows that the matrix $\posteriorcov$ consists of the model matrix $\modelmatrix$ and the {\priorcovariance} matrix $\vec{C}$. Both matrices depend on the choice of the training set. Straightforwardly, we could evaluate $\det(\posteriorcov)$ for all possible training sets. This brute force approach would yield the global minimum of $\det(\posteriorcov)$, but poses a NP-hard problem\cite{welch_algorithmic_complexity}. Therefore, we use Fedorov's algorithm\cite{10.2307/2346165} to approximate a D-optimal training set. Fedorov's algorithm works in the following way: First we randomly select an initial training set with the desired number of {\configurations}. Then, the algorithm iteratively tries to swap each {\configuration} from the training set with every {\configuration} from the pool of all {\configurations}. A swap is accepted if it decreases the determinant of the covariance matrix $\det{(\posteriorcov)}$. Once the algorithm has tried all {\configurations} from the pool and found the best swap, it proceeds to the next {\configuration} from the training set. This process repeats itself until it no longer finds a better training set, i.e. could not swap any {\configuration}. Fedorov's algorithm allows limiting the computational cost of the selection while retaining SAMPLE's quasi-deterministic nature.

 \subsection{D-optimal Selection Versus Random Selection}

\begin{figure}[H]
\includegraphics[width=1\linewidth]{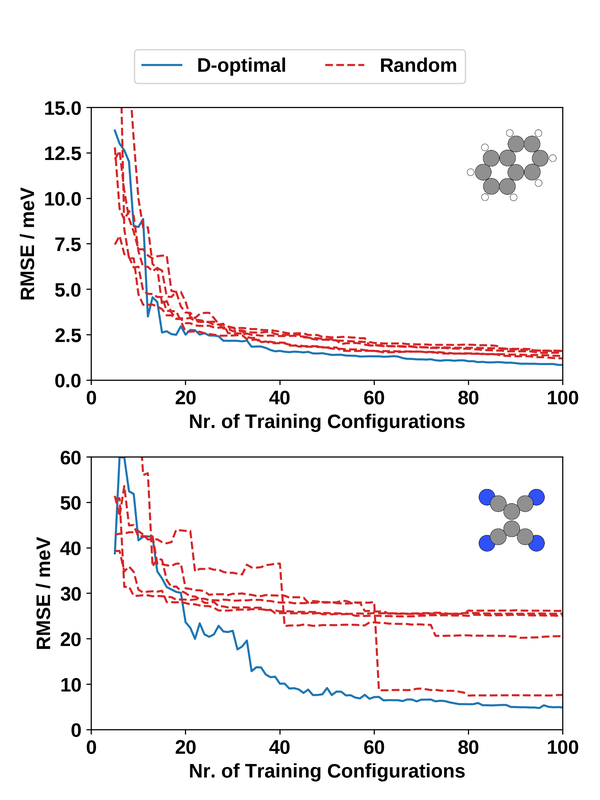}
\caption{Dependence of the root mean square error of the predicted adsorption energies per molecule on the number of training {\configurations} for D-optimal and random selection, (Top) Naphthalene test system, (Bottom) Tetracyanoethylene test system}
\label{fg:random_vs_D-optimal}
\end{figure}

The simplest way to generate an unbiased set of training data is to randomly select training points. But this approach suffers from a lack of reproducibility. One training set might describe the system well, while another one performs poorly (see figure \ref{fg:random_vs_D-optimal}). Hence, several SAMPLE runs, with different random training sets, would be required to generate a reliable prediction. D-optimal selection solves this problem, since the minimum determinant $\min \{ \det{(\posteriorcov)} \}$ is well defined. We demonstrate the advantage of D-optimal selection over random selection using two test systems: One for naphthalene on Cu(111), consisting of about $3200$ {\configurations} with $1$ to $6$ molecules per unit cell and one tetracyanoethylene on Cu(111) with about $1000$ {\configurations} with $1$ to $2$ upright standing molecules per unit cell. To reduce the computational effort, we strip the {\configurations} of their metal substrates, i.e. we only consider naphthalene/tetracyanoethylene sheets in vacuum. This is justified for demonstrative purposes, since the symmetries of the system and therefore the correlation between feature distances and energy differences are retained. Figure \ref{fg:random_vs_D-optimal} illustrates the improvement D-optimal selection provides over random selection. For relevant training set sizes ($N_{set} > 60$) D-optimal selection performs consistently better than random selection. Further, random training set selection leads to differently performing training sets, while D-optimality produces consistent results.

\subsection{Testing the Bayesian Linear Regression Algorithm}
\label{ssc:testing_BLR}

\begin{figure}[H]
\includegraphics[width=1\linewidth]{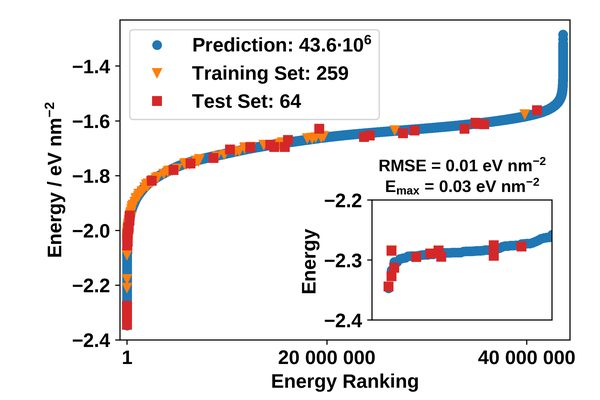}
\caption{Predicted adsorption energy of naphthalene on Cu(111), training set: $259$ D-optimally selected {\configurations}, test set: $64$ randomly selected {\configurations}}
\label{fg:S_curve}
\end{figure}

To test the Bayesian linear regression algorithm, we predict the adsorption energies of about $43.6\cdot 10^6$ {\configurations} as we have discussed in chapter \ref{ssc:configurations_for_naphthalene_on_cu111}. To determine the energy model we D-optimally select a training set of $259$ {\configurations} and calculate the adsorption energy with dispersion corrected DFT (PBE + TS\textsuperscript{surf})\cite{aims, PBE, TS-surf}. Using this model we are able to predict the adsorption energies of all {\configurations} generated by the procedure discussed in chapter \ref{ssc:configurations_for_naphthalene_on_cu111}. To test the prediction, we randomly select $64$ low energy {\configurations}, which were not part of the training set, and calculate their adsorption energy with DFT. Comparing the latter with the predicted adsorption energies allows us to determine a RMSE. This measure acts as a quality criterion for our prediction. Figure \ref{fg:S_curve} shows that there is excellent agreement between predicted and the $64$ calculated adsorption energies from the test set.

We find a root mean square error of $0.01~eV nm^{-2}$, which is smaller than the {\numericalDFTuncertainty} (about $0.05~eV nm^{-2}$). Additionally, this error is smaller than the often cited thermal energy scale $1~k_B T$, which corresponds to about $0.046~eV nm^{-2}$ for a coverage of $1.78~N_{ads} nm^{-2}$ ($10$ Cu atoms per molecule).

\section{Predicting Phase Diagrams}
\label{sc:predicting_phase_diagrams}

So far, we have discussed finding the coarse-grained potential energy surface using data from DFT calculations. By minimizing the adsorption energy, we can determine the most stable surface polymorph at $0~K$. For experiments and technical applications, however, the interesting temperature range lies significantly above absolute zero. In addition, materials can form different polymorphs at different temperatures and pressures. These polymorphs can have vastly different physical properties and only specific polymorphs of a material may be suitable for a given application.

In thermodynamic equilibrium a closed system (that can exchange work and heat, but not matter with its surroundings) seeks to minimize the Gibbs free energy. At an interface the measure of interest is therefore the Gibbs free energy of adsorption $\gamma$. Therefore, the polymorph that forms at a certain temperature and pressure is the one with the lowest $\gamma$. To determine the Gibbs free energy of adsorption $\gamma$, we employ {\abinitiothermodynamics}\cite{rogal_reuter}. The Gibbs free energy $\gamma$ is given by equation \ref{eq:gibbs_energy}. \hilight{As commonly done in literature\cite{rogal_reuter, doi:10.1002/adma.201404187, PhysRevB.65.035406} we neglect the contributions of the vibration enthalpy, the configuration entropy and the mechanical work, which leads to:

\begin{equation}
\gamma ~=~ \frac{1}{A} ( E - \mu_{ads} N_{ads})
\label{eq:gibbs_energy}
\end{equation}
}

\hilight{
Here, $E$ is the adsorption energy (the sum of {\oneandtwobodyinteractions}, see equation \ref{eq:energy_model_3}) as provided by SAMPLE. Further, $N_{ads}$ is the number of adsorbates in the unit cell and $\mu_{ads}$ is the chemical potential of the adsorbate in the gas phase. The latter can be calculated by using the following approximation\cite{rogal_reuter, cramer}:}

\begin{equation}
\mu_{ads} ~=~ \mu^{trans} + \mu^{rot}
\end{equation}

The chemical potential of the adsorbate in the gas phase $\mu_{ads}$ consists of $\mu^{trans}$, the translational and $\mu^{rot}$, the rotational chemical potential. We neglect vibrational, electronic and nuclear contributions to $\mu_{ads}$. First, consistency requires omitting the vibrational chemical potential, since we do not include the vibrational enthalpy for adsorbed molecules. Secondly, in most molecules electronic and nuclear excitation energies are large compared to $1~k_B T$. Hence, only the ground states, which we calculate with DFT, appreciably contribute to the chemical potential.\cite{rogal_reuter} The equations for the translational $\mu^{trans}$ and rotational chemical potential $\mu^{rot}$ can be found in various textbooks such as the one by Cramer\cite{cramer}.

\begin{equation}
\mu^{trans}(p,T) ~=~ -k_B T \ln \left( \frac{(k_B T)^{5/2} ( 2 \pi m)^{3/2} }{p h^3} \right)
\end{equation}

The translational chemical potential $\mu^{trans}$ is a function of temperature $T$ and pressure $p$. $m$ is the mass of the molecule.

\begin{equation}
\mu^{rot}(T) ~=~ -k_B T \ln \left[ \frac{\sqrt{\pi I_A I_B I_C}}{\sigma} \frac{(8 \pi^2 k_B T)^{3/2}}{h^3} \right]
\end{equation}

The rotational chemical potential $\mu^{rot}$, given here for a nonlinear molecule, is a function of temperature $T$. $I_A, I_B, I_C$ are the different moments of inertia and $\sigma$ is the symmetry number of the molecule. The latter is the number of pure rotations that map the molecule onto itself.

The chemical potential is a property of the molecules in the gas phase and only depends on its temperature and pressure, making it the same for all {\configurations}. Additionally, $\gamma$ depends on the adsorption energy $E$ and the area $A$ of the unit cell (or the coverage). Hence, for a particular coverage, only the {\configuration} with the lowest adsorption energy can show up in the phase diagram. This circumstance renders evaluating the Gibbs free energy of adsorption computationally inexpensive. This allows us to identify the {\configuration} with the smallest $\gamma$ for a range of temperatures and pressures and thereby generate a phase diagram.

\subsection{Phase Diagram for Naphthalene on Cu(111)}

\begin{figure}[h]
\includegraphics[width=1\linewidth]{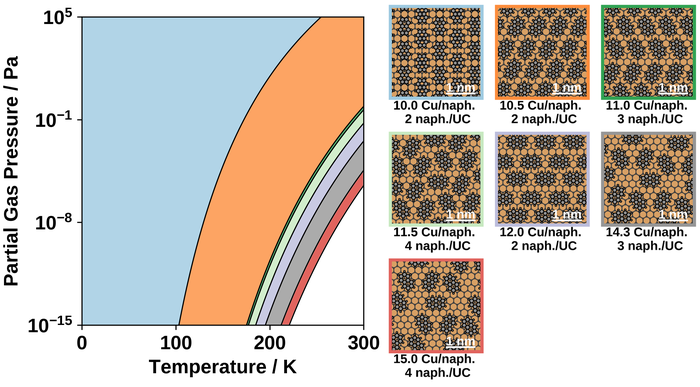}
\caption{Phase diagram for naphthalene on Cu(111)}
\label{fg:phase_diargam}
\end{figure}

\begin{figure}[h]
\includegraphics[width=1\linewidth]{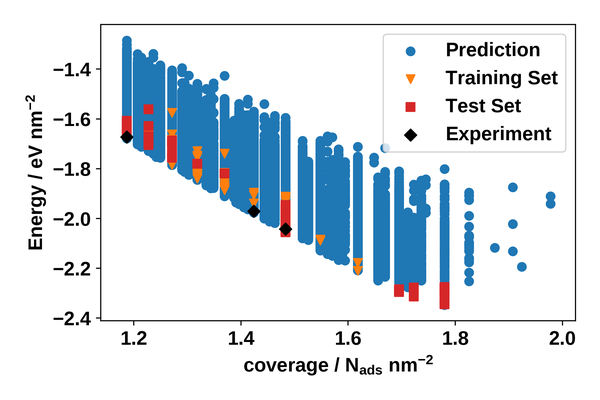}
\caption{Plot of the adsorption energies against their individual coverages}
\label{fg:prediction_over_coverage}
\end{figure}

We calculate the phase diagram for approximately $43.6\cdot 10^6$ {\configurations} as we have discussed in chapter \ref{ssc:configurations_for_naphthalene_on_cu111}. As depicted in figure \ref{fg:phase_diargam}, we observe seven different phases with a coverage ranging from  $1.19~N_{ads} nm^{-2}$ to $1.78~N_{ads} nm^{-2}$ ($15~A_{PUC}$ to $10~A_{PUC}$ naphthalene molecule) and $2$ thru $4$ molecules per unit cell. The energetically most favorable {\configuration} has $2$ molecules per unit cell and a coverage of $1.78~N_{ads} nm^{-2}$ ($10~A_{PUC}$ per naphthalene molecule). This {\configuration} also forms the phase with the highest packing density. {\configurations} with higher coverage become less favorable in energy, due to the repulsion of the naphthalene molecules (see figure \ref{fg:prediction_over_coverage}). The coverage decreases with decreasing pressure and increasing temperature. Further, if the temperature rises above roughly $300 ~K$ and the partial pressure of naphthalene drops roughly below $10^{-15} ~Pa$, it is no longer energetically favorable for naphthalene molecules to adsorb in the studied range of coverages.

In addition to the phase diagram, it is interesting to plot the adsorption energies of our $43.6\cdot 10^6$ {\configurations} against their individual coverages. Such an illustration, as depicted in figure \ref{fg:prediction_over_coverage}, yields a quantitative comparison of all {\configurations}.

\hilight{In figure \ref{fg:prediction_over_coverage} the {\configurations} with the lowest Gibbs free energies (at different temperatures and pressures) form a lower contour.} In principle, all {\configurations} within the energy uncertainty above this contour constitute equally valid phases. Therefore, it is interesting to take a look at these {\configurations} and figure out where the experimental {\configurations} can be found in the prediction. Table \ref{tb:ranking_of_experiment} illustrates the experimental phases. We find that the predicted energies of the experimental {\configurations} lie close to the lower contour, less than the {\numericalDFTuncertainty} (about $0.05~eV nm^{-2}$) above the predicted lowest-energy {\configurations}. Further we observe that the best adsorption energies per area decrease with increasing coverage until a coverage of $1.78~N_{ads} nm^{-2}$. Beyond this coverage, the best adsorption energies per area increase again.

Two experimentally determined phases have coverages that show up in the phase diagram, namely $1.48~N_{ads} nm^{-2}$ and $1.19~N_{ads} nm^{-2}$ ($12~A_{PUC}$ and $15~A_{PUC}$ per naphthalene molecule). Coverage $1.48~N_{ads} nm^{-2}$ corresponds to experimental phase (a) while coverage $1.19~N_{ads} nm^{-2}$ matches experimental phase (c) (see table \ref{tb:ranking_of_experiment}). Both experimental phases contain $1$ molecule per unit cell. A comparison between predicted and experimental {\configurations} yields the following results:

\begin{itemize}
\item Experimental phase (a) matches a {\configuration} whose predicted adsorption energy lies $(0.011 ~\pm~ 0.050) ~eV nm^{-2}$ higher than that of the best predicted {\configuration} of equal coverage. The experimental {\configuration} has rank $256$ among {\configurations} with coverage $1.48~N_{ads} nm^{-2}$.

\item Experimental phase (b) has a coverage of $1.42~N_{ads} nm^{-2}$ ($12.5~A_{PUC}$ per naphthalene molecule) and $6$ molecules per unit cell. We find a matching {\configuration} with an adsorption energy of $(0.005 ~\pm~ 0.050) ~eV nm^{-2}$ above the energetically most favorable {\configuration} of the same coverage. Hence, the experimental {\configuration} has rank $56$ among {\configurations} with coverage $1.42~N_{ads} nm^{-2}$.

\item Experimental phase (c) matches the second best predicted {\configuration} of equal coverage. The predicted adsorption energy of phase (c) is only negligibly higher ($0.0002 ~eV nm^{-2}$) than that of the best predicted {\configuration}. 
\end{itemize}

In all three cases the discrepancy is smaller than the {\numericalDFTuncertainty} (about $0.05~eV nm^{-2}$). Further, the molecules adopt the fcc-hollow {\localgeometry} in all matching {\configurations}. This is the energetically most favorable {\localgeometry}. Table \ref{tb:ranking_of_experiment} shows a summarized comparison between prediction and experiment. 


\begin{table}[h]
\setlength\extrarowheight{0pt}
\setlength{\tabcolsep}{2pt}
\small
\begin{tabular}{c !{\color{white}\vrule width 1mm} c !{\color{white}\vrule width 1mm} c !{\color{white}\vrule width 1mm} c}
\rowcolor{lightgray}
\parbox[c]{2.5cm}{\centering \textbf{Experiment} \cite{forker, yamada}} &
\parbox[][1.3cm][c]{2.5cm}{\centering \textbf{Best Pre- \\ diction per \\ Coverage}} &
\parbox[c]{1.1cm}{\centering $\Delta E ~/$ \\ $eV nm^{-2}$} &
\parbox[c]{0.8cm}{\centering \textbf{Rank \\ of}} \\
\myhline
\rowcolor{lightgray}
\parbox{2.45cm}{\vspace{2pt}\includegraphics[width=1\linewidth]{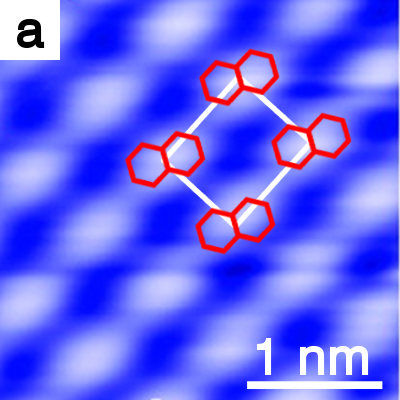}\\
$1.48~N_{ads} nm^{-2}$ \\ 1 M/UC \\ T=120 K \vspace{2pt}} &
\parbox{2.45cm}{\vspace{2pt}\includegraphics[width=1\linewidth]{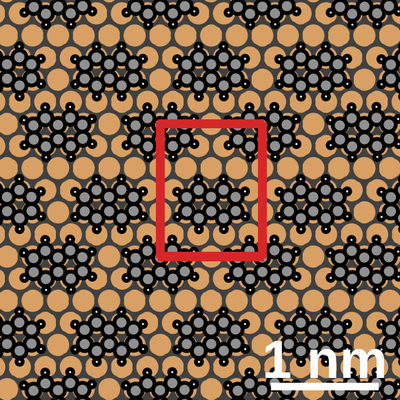}\\
$1.48~N_{ads} nm^{-2}$ \\ 2 M/UC \\ \vspace{2pt}} &
$0.011$ &
257 \\
\myhline
\rowcolor{lightgray}
\parbox{2.45cm}{\vspace{2pt}\includegraphics[width=1\linewidth]{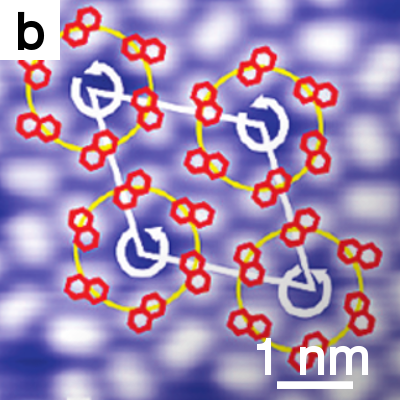}\\ 
$1.42~N_{ads} nm^{-2}$ \\ 6 M/UC \\ T=120 K \vspace{2pt}} &
\parbox{2.45cm}{\vspace{2pt}\includegraphics[width=1\linewidth]{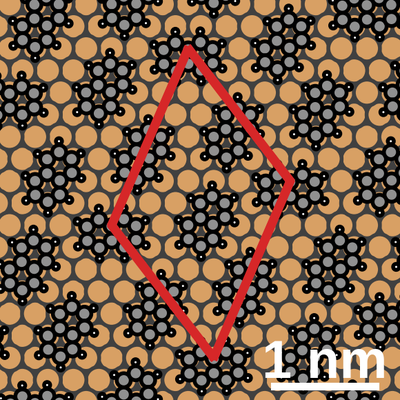}\\
$1.42~N_{ads} nm^{-2}$ \\ 4 M/UC \\ \vspace{2pt}} &
$0.005$ &
56 \\
\myhline
\rowcolor{lightgray}
\parbox{2.45cm}{\vspace{2pt}\includegraphics[width=1\linewidth]{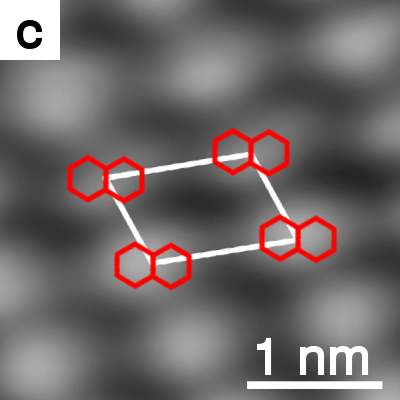}\\
$1.19~N_{ads} nm^{-2}$ \\ 1 M/UC \\ T=140 K \vspace{2pt}} &
\parbox{2.45cm}{\vspace{2pt}\includegraphics[width=1\linewidth]{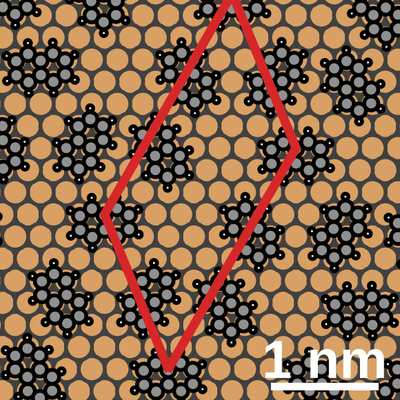}\\
$1.19~N_{ads} nm^{-2}$ \\ 4 M/UC \\ \vspace{2pt}} &
$0.0002$ &
2 \\
\end{tabular}

\caption{Ranking of the experimental phases in out prediction: $\Delta E$ is the energy difference between experimental phases and lowest energy {\configuration} of similar coverage, experimental figures modified with permission from \cite{forker, yamada}}
\label{tb:ranking_of_experiment}

\end{table}

\section{Conclusion}
\label{sc:conclusion}

In this publication we present SAMPLE, a comprehensive tool for surface structure search. SAMPLE is able to push back the configurational explosion by employing a coarse-grained model consisting of {\localgeometries} and unit cells. Furthermore, it makes use of optimal design theory in conjunction with a smart-data machine learning approach based on Bayesian linear regression. This allows for a quasi-deterministic exploration of the coarse-grained potential energy surface. SAMPLE requires only a small number of DFT calculations, serving as training data, to predict adsorption energies of millions of {\configurations}. This capability gives SAMPLE a competitive edge over stochastic structure search methods. Beyond that, evaluating the prediction results with {\abinitiothermodynamics} provides further insight into a system's polymorphism.

We demonstrate these capabilities using the example of naphthalene on Cu(111) and show SAMPLE's ability to efficiently predict adsorption and Gibbs free energies of millions of possible commensurate {\configurations} of molecules on surfaces. Moreover, the comparison with experiment demonstrates SAMPLE's predictive power, since all experimental polymorphs rank within the {\numericalDFTuncertainty} compared to the best prediction. Additionally, the phase diagram for naphthalene on Cu(111) yields an overview of the system's behavior as function of the deposition conditions. SAMPLE achieves these results from first principles and without input of experimental parameters.

\section{Acknowledgements}
\label{sc:acknowledgements}

Financial support by Austrian Science Fund (FWF) project P28631-N36 and the START award Y 1157-N36 is gratefully acknowledged. Computational results have been achieved in part using the Vienna Scientific Cluster (VSC). An award of computer time was provided by the Innovative and Novel Computational Impact on Theory and Experiment (INCITE) program. This research used resources of the Argonne Leadership Computing Facility, which is a DOE Office of Science User Facility supported under Contract DE-AC02-06CH11357.





\bibliographystyle{model1-num-names}
\bibliography{references}







\newpage
\section*{Glossary}

\begin{itemize}
	\setlength{\itemindent}{-0.55cm}
	
	\item[] $A_{PUC}$ \\ primitive substrate unit cell area
	
	\item[] {\configuration} \\ arrangement of {\localgeometries} in a unit cell
	
	\item[] {\configurationhash} \\ unique representation of a {\configuration} as a tuple of integer numbers
	
	\item[] {\decaypower}, $n$ \\ hyperparameter representing the exponent of a feature vector element
	
	\item[] {\distancecutoff}, $d^{AB}_{max}$ \\ maximal distance between respective atom-species of two molecules, beyond which they become effectively non-interacting
	
	\item[] {\distancethreshold}, $d^{AB}_0$ \\ minimal distance between respective atom-species of two molecules, below which strong Pauli repulsion occurs
	
	\item[] {\DFTuncertainty}, $\sigma_{model}$ \\ inherent uncertainty between the energy model and the DFT calculations, that remains even if the model is trained on an infinite number of DFT calculations
	
	\item[] $E$ \\ adsorption energy of a {\configuration}
	
	\item[] {\featurecorrelationlength}, $\xi$ \\ hyperparameter influencing the similarity measure between two \featurevectors
	
	\item[] {\featuredimension} \\ number of elements in the feature vector
	
	\item[] {\featurethreshold}, $\Delta f$ \\ distance in feature space that two {\featurevectors} need to have in order to be distinct
	
	\item[] {\featurevector}, $\vec{f_i}$  \\ representation of a pair of {\localgeometries} for Bayesian linear regression
	
	\item[] {\fractionalcoordinates} \\ coordinates in fractions of the primitive substrate lattice vectors
	
	\item[] $\gamma$ \\ Gibbs free energy of adsorption per area
	
	\item[] {\interactionvector} $\vec{\omega}$ \\ vector containing all {\oneandtwobodyinteractions}
	
	\item[] $\mathbf{l_i}$ \\ unit cell lattice vector
	
	\item[] {\likelihood} \\ likelihood as defined in Bayes' theorem; here probability for the adsorption energies given the {\oneandtwobodyinteractions}
	
	\item[] {\localadsorptiongeometry} \\ adsorption geometry of an isolated molecule on the substrate
	
	\item[] {\localgeometry} \\ {\localadsorptiongeometry} or one of its symmetry-equivalents
	
	\item[] $\mathbf{M}$ \\ epitaxy matrix
	
	\item[] $\mu_{ads}$ \\ chemical potential of the molecules in gas phase
	
	\item[] {\modelmatrixtext}, $\vec{X}$ \\ matrix consisting of several stacked {\modelvectors} $\vec{n}$
	
	\item[] {\modelvector}, $\vec{n}$ \\ vector encoding which {\oneandtwobodyinteractions} occur in a {\configuration}
	
	\item[] $N_{ads}$ \\ number of adsorbates per unit cell
	
	\item[] {\textit{\numericalDFTuncertainty}} \\ uncertainty derived from the convergence of the DFT settings
	
	\item[] {\onebodyinteraction}, $U_i$ \\ adsorption energy of a single {\localgeometry}
	
	\item[] {\onebodyinteractionuncertainty}, $\sigma_{one-body}$ \\ standard deviation of the prior for {\onebodyinteraction}
	
	\item[] {\posterior} \\ posterior probability as defined in Bayes' theorem; here probability for the {\oneandtwobodyinteractions} given the adsorption energies 
	
	\item[] {\posteriorcovariance}, $\posteriorcov$ \\ covariance matrix of the Gaussian representing the posterior probability
	
	\item[] {\posteriormean}, $\bar{\vec{\omega}}$ \\ mean value of the Gaussian representing the posterior probability, best prediction for the {\oneandtwobodyinteractions}
	
	\item[] {\prior} \\  prior probability as defined in Bayes' theorem; here prior knowledge about the parameters of the energy model
	
	\item[] {\priorcovariance}, $\vec{C}$ \\ covariance matrix of the Gaussian representing the prior probability; prior knowledge about the correlations of different {\oneandtwobodyinteractions} in the energy model
	
	\item[] {\priormean}, $\vec{\omega_0}$ \\ mean value of the Gaussian representing the prior probability; prior knowledge about the {\oneandtwobodyinteractions} in the energy model
	
	\item[] {\realspacedecaylength}, $\tau$ \\ hyperparameter controlling how quickly interactions decay with distance
	
	\item[] {\standardunitcell} \\ most compact representation of a unit cell
	
	\item[] $\vec{T}$ \\ transformation matrix
	
	\item[] {\twobodyinteraction}, $V_p$ \\ interaction energy between a pair of {\localgeometries}
	
	\item[] {\twobodyinteractionuncertainty}, $\sigma_{two-body}$ \\ standard deviation of the prior for {\twobodyinteractions}
	
	\item[] $\mathbf{v_i}$ \\ primitive substrate lattice vector
	
\end{itemize}

\end{document}


\begin{frontmatter}



\title{Supplementary Information}

\author{Lukas H\"ormann, Andreas Jeindl, Alexander T. Egger, Michael Scherbela and Oliver T. Hofmann}

\address{Institute of Solid State Physics, NAWI Graz, Graz University of Technology, Petersgasse 16, 8010 Graz, Austria}

\end{frontmatter}

\hilight{
\section{Epitaxy Matrix and Types of Coincidence}

In general, we can represent the unit cell of any ordered adsorbate layer in fractional coordinates of the primitive substrate lattice vectors ({\fractionalcoordinates}). The transformation between Cartesian and {\fractionalcoordinates} can be accomplished with an epitaxy matrix $\mathbf{M}$.

\begin{equation}
\begin{pmatrix} \mathbf{l_1} \\ \mathbf{l_2} \\ \end{pmatrix} = \mathbf{M} \cdot \begin{pmatrix} \mathbf{v_1} \\ \mathbf{v_2} \\ \end{pmatrix} =  \begin{pmatrix} m_1 & m_2 \\ m_3 & m_4 \\ \end{pmatrix} \cdot \begin{pmatrix} \mathbf{v_1} \\ \mathbf{v_2} \\ \end{pmatrix}
\label{eq:epitaxy}
\end{equation}

Here $\mathbf{l_1}$ and $\mathbf{l_2}$ are the two-dimensional super-lattice vectors of the adsorbate unit cell, $\mathbf{v_1}$ and $\mathbf{v_2}$ are the primitive substrate lattice vectors and $\mathbf{M}$ is the epitaxy matrix. Several types of epitaxy exist (see figure \ref{fg:types_of_epitaxy})\cite{epitaxy}:

\begin{itemize}
	\item[(a)] commensurability: All elements of the epitaxy matrix are integers $m_i \in \mathbb{Z}$ (see equation \ref{eq:epitaxy}). 
	
	\item[(b)] point-on-line coincidence: The adsorbate lattice lies on lines corresponding to the primitive substrate lattice.
	
	\item[(c)] coincidence II (or higher order commensurability): Only a fraction of the adsorbate lattice points coincide with the substrate, i.e. only the points of an adsorbate lattice supercell coincide.
	
	\item[(d)] incommensurability: The adsorbate lattice does not coincide with the substrate lattice or lattice lines thereof.
\end{itemize}

\begin{figure}[H]
	\setlength{\tabcolsep}{2pt}
	\begin{tabular}{cccc}
		\includegraphics[width=0.235\linewidth]{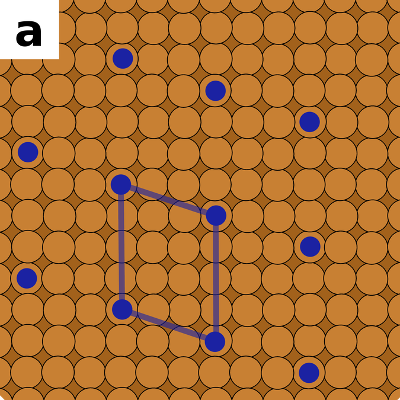} &
		\includegraphics[width=0.235\linewidth]{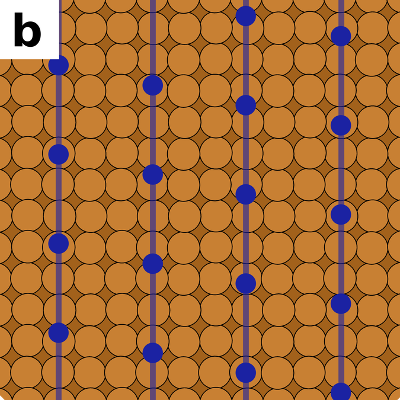} &
		\includegraphics[width=0.235\linewidth]{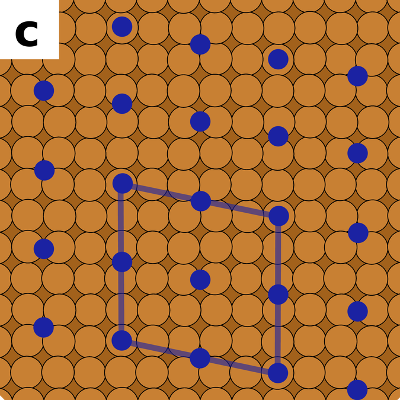} &
		\includegraphics[width=0.235\linewidth]{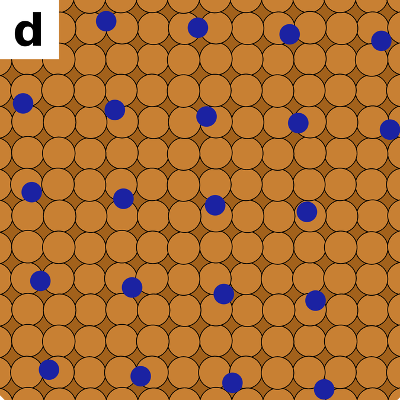}
	\end{tabular}
	\caption{Types of epitaxy: (a) commensurability, (b) point-on-line coincidence, (c) coincidence II, (d) incommensurability}
	\label{fg:types_of_epitaxy}
\end{figure}

\section{Geometric Considerations on Possible {\Configurations}}
\label{SI_sc:geometric_consideration_on_possible_configurations}

In commensurate {\configurations}, the smallest repeating unit contains only a small number of molecules. Even if we expand the definition to higher-order commensurate {\configurations}, the number of molecules per unit cell still remains small. For instance, commensurate {\configurations} of naphthalene on Cu(111) contain at most $6$ molecules per unit cell\cite{forker, yamada}. Hence, it is sensible to limit the number of molecules per unit cell and thereby limit the number of possible {\configurations}.

Since {\closepacked} {\configurations} exhibit a large coverage $\Theta = N / A$ (number of molecules per area), we can enforce {\closepacking} by introducing a maximum area for the possible adsorbate unit cells for a given number of molecules. Similarly, the fact that at least one molecule must fit in the unit cell provides a minimum area for the adsorbate unit cell. To avoid very elongated unit cells, we use the width of the adsorbate molecule as a lower boundary for the width of the unit cell. We enforce this criterion by a minimum unit cell width that is defined as the height of the parallelogram, formed by the super-lattice vectors (see figures \ref{fg:unit_cell_parameters} and \ref{fg:narrow_unit_cell}).

\begin{figure}[H]
	\begin{minipage}[t]{0.48\linewidth}
		
		\includegraphics[width=0.66\linewidth]{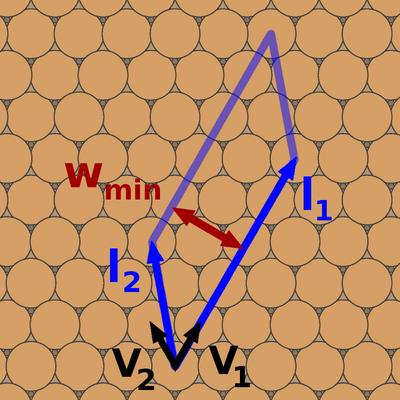}
		\caption{Parameters of a commensurate unit cell on a (111) surface: $\mathbf{l_1}$, $\mathbf{l_2}$ are the lattice vectors of the unit cell and $\mathbf{v_1}$, $\mathbf{v_2}$ are the primitive lattice vectors of the substrate, $w_{min}$ is the minimum width of the unit cell}
		\label{fg:unit_cell_parameters}
		
	\end{minipage}%
	\begin{minipage}[t]{0.04\linewidth}
		\hfill
	\end{minipage}%
	\begin{minipage}[t]{0.48\linewidth}
		
		\includegraphics[width=0.66\linewidth]{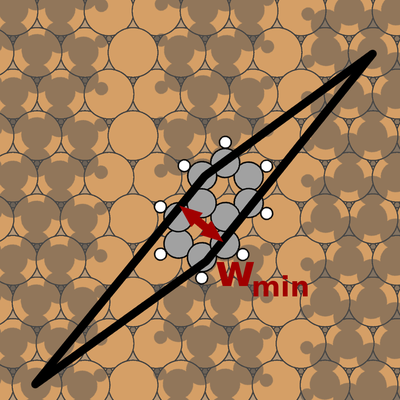}
		\caption{Example of an extremely elongated unit cell that is too narrow to accommodate a naphthalene molecule ($w_{min}$ is too small)}
		\label{fg:narrow_unit_cell}
		
	\end{minipage}
	
\end{figure}

Comprehensive structure search requires generating a set of all possible {\configurations} that fulfill the aforementioned considerations. Within the SAMPLE approach, a {\configuration} consists of a number of {\localadsorptiongeometries} placed in a unit cell. Hence, the procedure to generate {\configurations} necessitates comprehensive sets of unit cells as well as {\localadsorptiongeometries}. Hereafter we will elaborate on the concepts employed in generating such a comprehensive set of {\configurations}.
}

\section{Calculation of the Largest Element of the Epitaxy Matrix}
\label{SI_sc:calculation_of_the_largest_element_of_the_epitaxy_matrix}

We propose epitaxy matrices by iterating over different values of the epitaxy matrix elements $m_i$. Generating all unit cells necessitates choosing limits for this iteration. Finding the limits requires some geometric considerations. The range within which we need to vary the matrix elements $m_i$ depends on the largest super-lattice vector of a unit cell with a given area and a given minimum width. Using the area $A$ and the minimum width $w_{min}$ allows calculating the maximum length $\mathcal{l}_{max}$ as follows:
\begin{equation}
\mathcal{l}_{max} ~=~ \frac{A}{w_{min}}\\
\end{equation}

We can also calculate $\mathcal{l}_{max}$ by using the primitive lattice vectors $\mathbf{v_1}$ and $\mathbf{v_2}$, as well as the epitaxy matrix elements $m_1$ and $m_2$:
\begin{equation}
\mathcal{l}_{max}^2 ~=~ (\mathbf{v_1} m_1 + \mathbf{v_2} m_2)^2\\
\label{eq:max_length}
\end{equation}

We want to find the maximum values for $m_1$ and $m_2$. To simplify the notation we define the unit cell vectors $\mathcal{l}_1$ and $\mathcal{l}_2$ as well as the cosine $\cos{\alpha}$ of the angle $\alpha$ between them. 

\begin{tabular}{ccccc}
$\mathbf{\mathcal{l}_1} ~=~ \mathbf{v_1} m_1$
& &
$\mathbf{\mathcal{l}_2} ~=~ \mathbf{v_2} m_2$
& &
$\cos \alpha ~=~ \frac{\mathbf{v_1} \mathbf{v_2}}{|\mathbf{v_1}| |\mathbf{v_2}|}$
\end{tabular}

The law of cosines allows rewriting equation \ref{eq:max_length}.
\begin{align}
\mathcal{l}_{max}^2 ~&=~ \mathcal{l}_1^2 + \mathcal{l}_2^2 - 2 \mathcal{l}_1 \mathcal{l}_2 ~\cos \alpha
\end{align}

Solving this for $\mathcal{l}_1$ gives:
\begin{align}
\label{eq:l1(l2)}
\mathcal{l}_1(\mathcal{l}_2) ~&=~ \mathcal{l}_2 ~\cos \alpha \pm \sqrt{\mathcal{l}_2^2 ~\cos^2 \alpha - \mathcal{l}_2^2 + \mathcal{l}_{max}^2}
\end{align}

To find the maximum of $\mathcal{l}_1$, we need to take the derivative with respect to $\mathcal{l}_2$ and determine the roots.
\begin{align}
\frac{d \mathcal{l}_1}{d \mathcal{l}_2} ~&=~ \cos \alpha \pm \frac{1}{\sqrt{\cdots}} ~ \mathcal{l}_2 ~ (\cos^2 \alpha - 1) ~=~ 0
\end{align}

We solve the above relation for $\mathcal{l}_2$.
\begin{align}
\mathcal{l}_2 ~=~ \pm \left[ \frac{\mathcal{l}_{max}^2 ~ \cos^2 \alpha}{1 - \cos^2 \alpha} \right]^{1/2}
\label{eq:l2}
\end{align}

Substituting $\mathcal{l}_2$ in equation \ref{eq:l1(l2)} results in the expression for the maxima of $\mathcal{l}_1(\mathcal{l}_2)$. Equation \ref{eq:l2} yields two solutions for $\mathcal{l}_2$, a positive and a negative one. Hence, equation \ref{eq:l1(l2)} has four solutions. We are only interested in the maximum absolute value $\mathcal{l}_{1,max}$ of $\mathcal{l}_1(\mathcal{l}_2)$, which leads to the following expression. Due to symmetry, the solution for $\mathcal{l}_2(\mathcal{l}_1)$ is the same.
\begin{align}
\mathcal{l}_{1,max} ~&=~ |\mathcal{l}_2| ~|\cos \alpha| + \sqrt{\mathcal{l}_2^2 ~\cos^2 \alpha - \mathcal{l}_2^2 + \mathcal{l}_{max}^2}
\end{align}

Now we calculate the maximum value that $m_i$ can take round it to the next largest interger number.
\begin{align}
m_1 ~&=~ \ceil[\bigg]{ \frac{\mathcal{l}_{1,max}}{|\mathbf{v_1}|} }\\
m_2 ~&=~ \ceil[\bigg]{ \frac{\mathcal{l}_{1,max}}{|\mathbf{v_2}|} }\\
m_{max} ~&=~ max(m_1, m_2)
\end{align}

In order to find all unit cells, we have to vary the elements $m_i$ of the epitaxy matrix within a range given by $m_{max}$:
\begin{equation}
m_i \in [-m_{max}, m_{max}]
\end{equation}

\section{Algorithmic Details for the Standard Unit Cell}

In principle, for every unit cell an infinite number of equivalent cells exist. However, by defining a set of conclusive criteria it is possible to select one of these equivalent unit cells as the standard unit cell. Transforming a unit cell into a standard unit cell requires two types of transformations: First, combinations of lattice vectors allow generating more compact unit cells and secondly, symmetry transformations enable orienting the unit cell such that it better fulfills the criteria for the standard unit cell.

\subsection{Criteria for the \textit{Standard Unit Cell}}
\label{SI_ssc:criteria_for_the_standard_unit_cell}

To find a conclusive \textit{standard unit cell} we require a number of criteria for the epitaxy matrix. These criteria are defined in hierarchical order, with criteria higher up in the hierarchy trumping the ones with lower priority. This avoids conflicts between criteria.

First, we minimize the larger diagonal of the unit cell. This criterion enforces compact unit cells.
\begin{itemize}
\item[1.] $min(d_{max}^2)$
\end{itemize}

Secondly, we compare elements of the epitaxy matrix. Criteria 2 thru 5 enforce that the epitaxy matrix is as close to a diagonal matrix as possible, i.e. the absolute value of the diagonal elements should be large, that of the off-diagonal elements should be small.
\begin{itemize}
\item[2.] $|m_0| \geqslant |m_1|$
\item[3.] $|m_3| \geqslant |m_2|$
\item[4.] $|m_0| \geqslant |m_3|$
\item[5.] $|m_2| \geqslant |m_1|$
\end{itemize}

Thirdly, elements of the epitaxy matrix should be positive. These criteria favor unit cells that lie in the first quadrant.
\begin{itemize}
\item[6.] $m_0 \geqslant 0$
\item[7.] $m_1 \geqslant 0$
\item[8.] $m_2 \geqslant 0$
\item[9.] $m_3 \geqslant 0$
\end{itemize}

If two unit cells fulfill criteria 1 thru 9, we pick the unit cell with the larger first element in the epitaxy matrix.
\begin{itemize}
\item[10.] $max(m_1)$
\end{itemize}

\subsection{Parameters in the Transformation Matrix for the Standard Unit Cell}
\label{SI_ssc:parameters_in_the_transformation_matrix_for_the_standard_unit_cell}

The first type of transformation, the combination of lattice vectors, is equivalent to a transformation of the epitaxy matrix. Hereby, equivalent unit cells can be obtained by using the transformation shown in equation \ref{eq:combination_of_lattice_vectors}, whereby the parameters of the transformation matrix $\vec{T}$ must fulfill $a, b, c, d \in \mathbb{Z}$ and $|\det{\vec{T}}| = 1$:
\begin{equation}
\begin{pmatrix} m_1' & m_2' \\ m_3' & m_4' \\ \end{pmatrix} ~=~ \underbrace{\begin{pmatrix} a & b \\ c & d \\ \end{pmatrix}}_{\vec{T}} \cdot \begin{pmatrix} m_1 & m_2 \\ m_3 & m_4 \\ \end{pmatrix}
\label{eq:combination_of_lattice_vectors}
\end{equation}

Finding useful limits for $a, b, c, d$ will reduce the computational effort for finding the standard unit cell. Let us first consider the parameters $a$ and $d$. If $b=c=0$, the parameters $a$ and $d$ combine a lattice vector with itself. Setting $a=1$ and $d=1$ yields the original unit cell. Setting $a=2$ or $d=2$ doubles the length of the respective lattice vector and thereby doubles the area of the unit cell. Larger values of $a$ and $d$ increase the area even further, hence the upper limit is $a, d \leqslant 1$ ($b=c=0$). Setting $a=-1$ or $d=-1$ subtracts the lattice vector twice from itself, reversing its direction. This is equivalent to a sign transformation. Hereby the area is conserved. However, larger negative values of $a$ and $d$ increase the area. Hence, the lower limit is $a, d \geqslant -1$. For a general case $b \neq 0$ and/or $c \neq 0$ this argument does not hold.

The parameters $b$ and $c$ control the shear of the transformed unit cell. If $a=d=1$, any possible values of either $b$ or $c$ alter the direction of the respective lattice vector, but conserve the area of the unit cell. Conversely, varying both $b$ and $c$ at the same time may not conserve the area.

To find limits for the iteration of $b$ and $c$, we remember that the standard unit cell should be as compact as possible. This means that its \textit{shear} should be minimal. Therefore, we can use the shear of the current unit cell to calculate the limits for the iteration of $b$ and $c$. We can calculate the shear as follows, whereby we round to the next larger integer since $a, b, c, d \in \mathbb{Z}$.
\begin{align*}
s_1 ~&=~ \left\lceil |\mathbf{l_1} \mathbf{l_2}| ~/~ |\mathbf{l_2}|^2 \right\rceil \\
s_2 ~&=~ \left\lceil |\mathbf{l_1} \mathbf{l_2}| ~/~ |\mathbf{l_1}|^2 \right\rceil \\
s_{max} ~&=~ max(s_1, s_2)
\end{align*}

To guarantee finding the standard unit cell, we vary b and c within a range given by $s_{max}$.
\begin{equation}
b, c \in [-s_{max}, s_{max}]
\end{equation}

\hilight{
\subsection{Algorithm to Find for the Standard Unit Cell}
\label{SI_ssc:implementation_of_the_algorithm}

In order to find the standard unit cell, we need to apply all combinations of $a, b, c, d$ that are within the limits outlined in chapter \ref{SI_ssc:criteria_for_the_standard_unit_cell} as well as all symmetry transformations included in the substrate's point-group. We can do so by iterating the parameters $a, b, c, d$ and applying symmetry transformations. This is done in five nested loops. The first four loops iterate $a, b, c, d$ respectively while the fifth loop performs the symmetry transformations. Hereby, we enforce a conservation of the area. Figure \ref{fg:combine_epitaxy_matrices} illustrates a number of examples for the transformation matrices $\vec{T}$ with the parameters $a, b, c, d$ and the resulting unit cells, while figure \ref{fg:rotate_unit_cell} shows the effect of symmetry transformations.

\begin{figure}[h]
	\setlength{\tabcolsep}{2pt}
	\begin{tabular}{ccc}
		\includegraphics[width=0.32\linewidth]{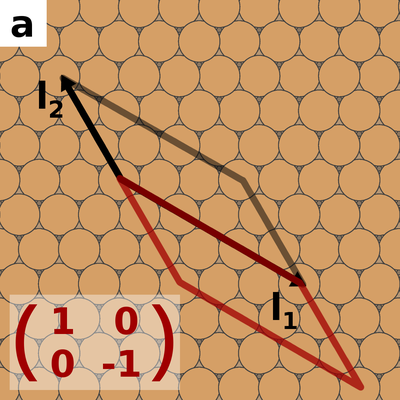} &
		\includegraphics[width=0.32\linewidth]{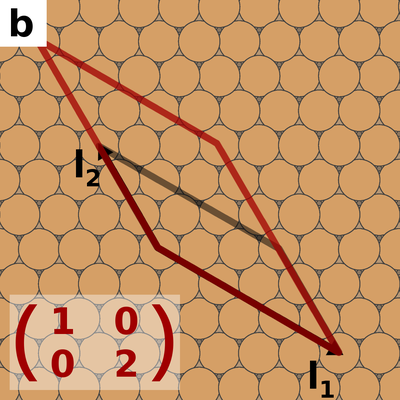} &
		\includegraphics[width=0.32\linewidth]{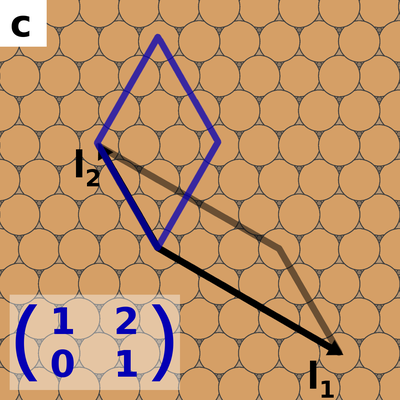} \\
		\includegraphics[width=0.32\linewidth]{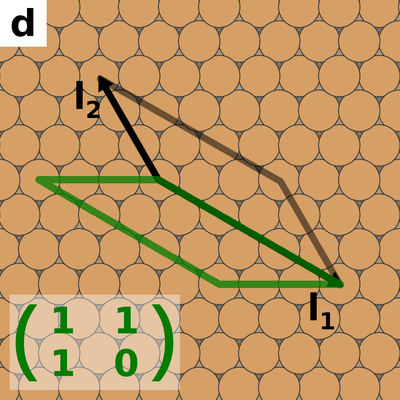} &
		\includegraphics[width=0.32\linewidth]{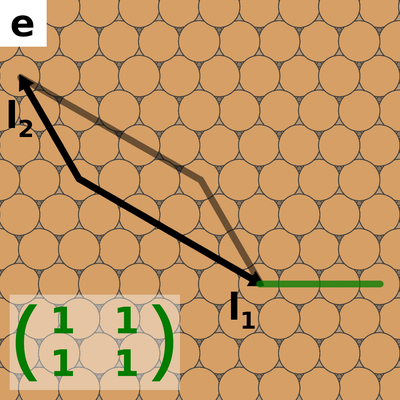} &
		\includegraphics[width=0.32\linewidth]{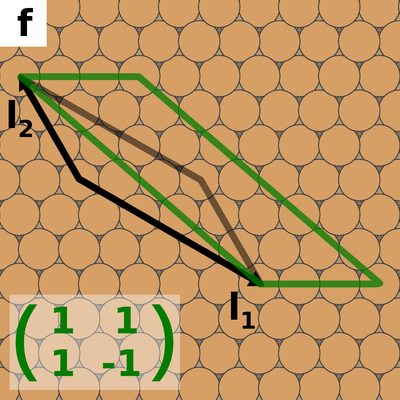}
	\end{tabular}
	\caption{Examples of combining lattice vectors using a transformation matrix $\vec{T}$: (a) transformation with $\det(\vec{T}) = -1$ conserves the area, (b) $\det(\vec{T}) = 2$ results in doubling the unit cell area and hence a non-equivalent unit cell, (c) shear through $b=2$ conserves the area, (d) general valid transformation with $det(\vec{T}) = -1$, (e) if $\det(\vec{T}) = 0$ the area becomes $0$, (f) example with $\det(\vec{T}) = 2$}
	\label{fg:combine_epitaxy_matrices}
\end{figure}

\begin{figure}[h]
	\setlength{\tabcolsep}{2pt}
	\begin{tabular}{ccc}
		\includegraphics[width=0.32\linewidth]{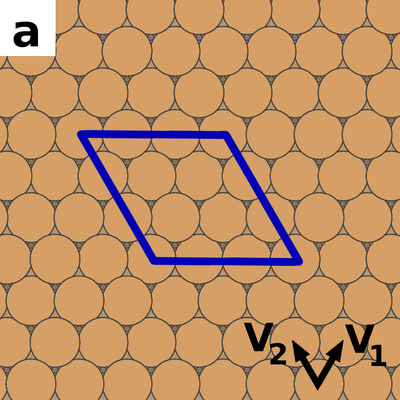} &
		\includegraphics[width=0.32\linewidth]{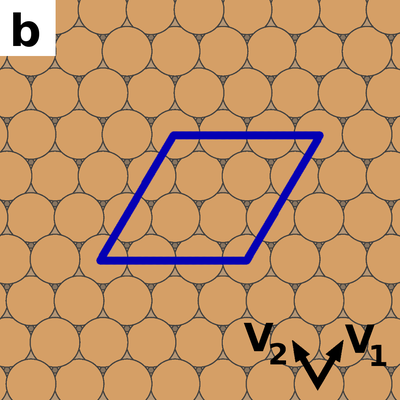} &
		\includegraphics[width=0.32\linewidth]{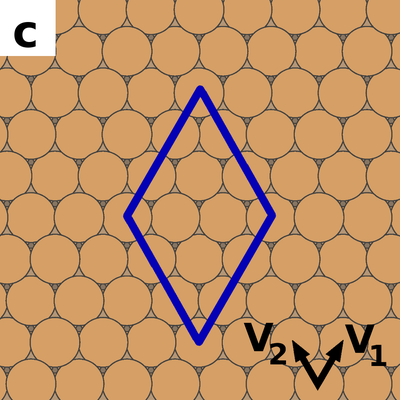} \\
		$\begin{pmatrix} 0 & 3 \\ -3 & 3 \\ \end{pmatrix}$ &
		$\begin{pmatrix} 3 & 0 \\ -3 & 3 \\ \end{pmatrix}$ &
		$\begin{pmatrix} 3 & 0 \\ 0 & 3 \\ \end{pmatrix}$
	\end{tabular}
	\caption{Examples of symmetry-equivalent unit cells: (a) epitaxy matrix with large non-diagonal elements, (b) epitaxy matrix fulfills criteria better than (a), (c) diagonal epitaxy matrix, which is the {\standardunitcell} for this cell-shape}
	\label{fg:rotate_unit_cell}
\end{figure}

After every transformation, the unit cell is evaluated regarding the criteria for a standard unit cell. If the algorithm finds a unit cell that better fulfills the criteria, the iteration stops and the process of combining lattice vectors and applying symmetry transformations starts anew. If a loop ends without finding a better unit cell, the current unit cell is the standard unit cell.

\section{Generating Unit Cells}
\label{SI_sc:generating_unit_cells}

In this chapter we aim to deliberate the technical aspects of generating unit cells according to the SAMPLE approach. As discussed in chapter \ref{SI_sc:geometric_consideration_on_possible_configurations}, we intend to find {\closepacked} commensurate {\configurations}. This goal entails two considerations:
\begin{itemize}
	\item The range of unit cell areas defines the possible coverages and number of molecule per unit cell.
	\item A minimum width of the unit cell should avoid generating narrow unit cells that cannot accommodate any molecules.
\end{itemize}

Usually, we are interested in unit cells of a given area. The area, given in units of the primitive substrate unit cell area $A_{PUC}$, directly relates to the epitaxy matrix via the following equation:
\begin{equation}
A = |m_1 m_4 - m_2 m_3| ~A_{PUC}
\label{eq:unit_cell_area}
\end{equation}

To generate possible unit cells, we iteratively propose different epitaxy matrices by varying the epitaxy matrix elements $m_i$. Using equation (\ref{eq:unit_cell_area}) we calculate the area of each proposed unit cell and only accept those with the desired area, i.e. we introduce an area constraint.

In order to find all possible unit cells, we require all combinations of the epitaxy matrix elements within a certain range $m_i \in [-m_{max}, m_{max}]$. A discussion on finding the maximal epitaxy matrix element $m_{max}$ can be found in the supplementary information.

To systematically generate all aforementioned combinations, we can iterate the epitaxy matrix elements $m_i$ with four nested loops, one for each element. Each loop starts the iteration at $m_i = -m_{max}$ and ends it at $m = m_{max}$. The computational effort of this task scales as $\mathcal{O}(m_{max}^4)$. However, using the area constraint reduces the number of independent epitaxy matrices elements $m_i$ to three. The fourth $m_i$ results automatically from equation (\ref{eq:unit_cell_area}). It is possible to reduce the computational effort to $\mathcal{O}(m_{max}^3)$, by considering three cases:

\begin{enumerate}
	\item \underline{$m_1 \neq 0$}\\
	Equation \ref{eq:unit_cell_area} allows calculating the epitaxy matrix element $m_4$ directly. 
	\begin{align}
	m_4 ~&=~ \frac{A + m_2 m_3}{m_1}
	\end{align}
	\item \underline{$m_1 = 0$ and $m_2 \neq 0$}\\
	In this case, the area is $A = m_2 m_3$ and $m_4$ becomes a free parameter. Hence, we calculate the value of $m_3$ and iterate $m_4$ within $[-m_{max}, m_{max}]$.
	\begin{align}
	m_3 ~&=~ A ~/~ m_2
	\end{align}
	\item \underline{$m_1 = 0$ and $m_2 = 0$}\\
	The area is always 0.
\end{enumerate}

\textbf{Unit cells for naphthalene on Cu(111).} For the example of naphthalene on Cu(111) the algorithm generates $987$ unique unit cells with areas between $0.562~nm^2$ ($10$ surface atoms) and $4.494~nm^2$ ($80$ surface atoms). We set the minimum width to $w_{min}=\length{4}$, slightly smaller than the naphthalene molecule. 
}

\section{Convergence Tests for Bayesian Linear Regression}
\label{SI_sc:BLR_Convergence_Test}
To converge settings for SAMPLE's Bayesian linear regression algorithm, we use a test system of $3222$ {\configurations} whose metal substrates have been removed. The training set comprises $222$ {\configurations}. For the test set we use all $3222$ {\configurations}.

\subsection{Feature Threshold}

\begin{figure}[h!]
	\includegraphics[width=\linewidth]{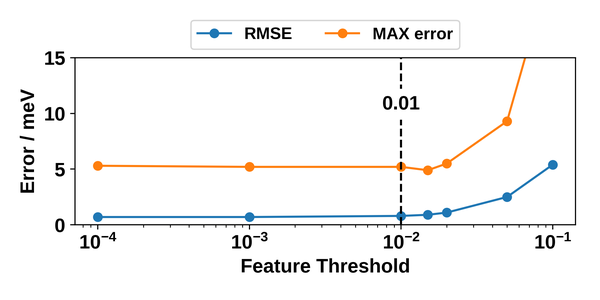}
	\caption{Convergence of the feature threshold}
	\label{fig:feature_threshold}
\end{figure}

We determine the {\featurethreshold} using the test system described in this chapter. We calculate the root mean square error per molecule as well as the maximal error by predicting all $3222$ {\configurations} and comparing the results to the DFT calculations. We find that the root mean square error converges for a {\featurethreshold} of $\Delta f = 0.01$ and below (see figure \ref{fig:feature_threshold}). Since a smaller {\featurethreshold} reduces the computational effort we use a {\featurethreshold} of $\Delta f = 0.01$.

\subsection{Feature Correlation Length}

\begin{figure}[h!]
	\includegraphics[width=\linewidth]{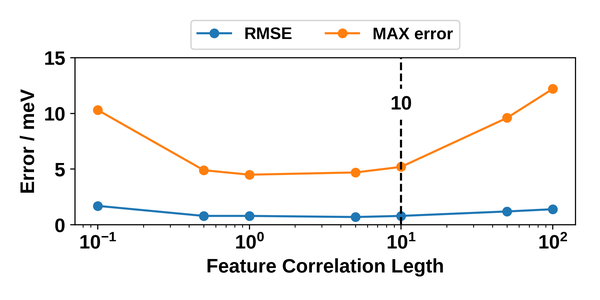}
	\caption{Convergence of the feature correlation length}
	\label{fig:feature_correlation_length}
\end{figure}

We determine the {\featurecorrelationlength} $\xi$ using the same procedure as for the {\featurethreshold}. We find that the root mean square error converges for a {\featurecorrelationlength} between $\xi = 0.5$ and $\xi = 10$ (see figure \ref{fig:feature_correlation_length}). We use a {\featurecorrelationlength} of $\xi = 10$.

\hilight{
\subsection{Real Space Decay Length}

We determine the {\realspacedecaylength} $\tau$ using the same procedure as for the {\featurethreshold}. We find that the root mean square error converges for a {\realspacedecaylength} above $\tau = 0.5~nm$. Note that a large {\realspacedecaylength} effectively sets the coefficients $\sigma_i^*$ to a constant.

\begin{figure}[h!]
	\includegraphics[width=\linewidth]{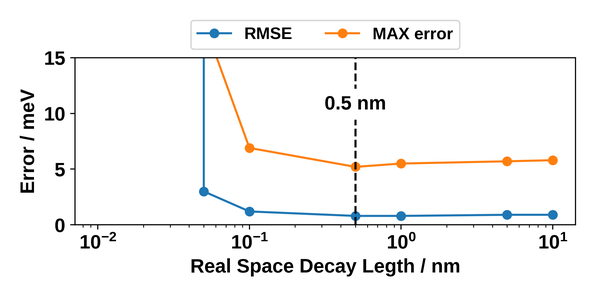}
	\caption{Convergence of the real space decay length}
	\label{fig:real_space_length}
\end{figure}

\subsection{Distance Cutoff}

We determine the {\distancecutoff} $d_{max}$ using the same procedure as for the {\featurethreshold}. The root mean square error converges for a {\distancecutoff} above $d_{max} = 1~nm$.
\begin{figure}[h!]
	\includegraphics[width=\linewidth]{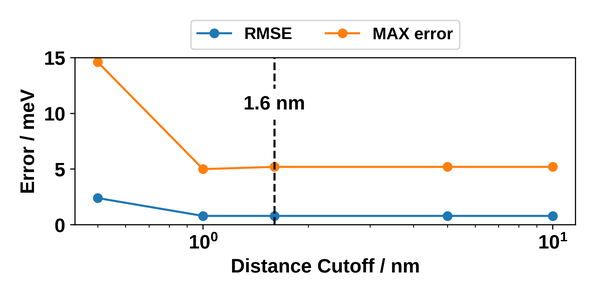}
	\caption{Convergence of the distance cutoff}
	\label{fig:distance_cutoff}
\end{figure}

\subsection{Feature Dimension}
\label{SI_ssc:feature_dimension}

We determine the {\featuredimension} using the same procedure as for the {\featurethreshold}. The root mean square error converges for a {\featuredimension} above $8$.

\begin{figure}[h!]
	\includegraphics[width=\linewidth]{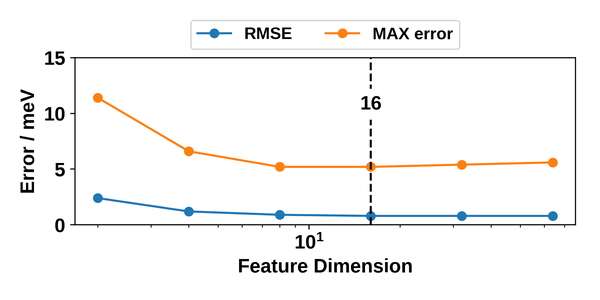}
	\caption{Convergence of the feature dimension}
	\label{fig:feature_dimension}
\end{figure}

\subsection{Dimensionality of the {\InteractionVector}}
\label{SI_ssc:dimensionality_of_the_interactionVector}

The dimensionality of the {\interactionvector} is controlled by the {\distancethreshold}, the {\distancecutoff}, the {\featurethreshold}, the {\decaypower} and the {\featuredimension}. Here, we change the dimensionality by setting the {\distancecutoff} to $d_{max} = 100~\AA$ while varying the {\featurethreshold}. Training and test set are the same as for the convergence of the {\featurethreshold}. The root mean square error converges for a dimensionality of about $400$ and above. We note that a large dimensionality significantly taxes computation resources, whereby the last data point in figure \ref{fig:number_of_features} requires about $15~min$ computation time (for the final parameter settings the computation time ranges in the seconds).

\begin{figure}[h!]
	\includegraphics[width=\linewidth]{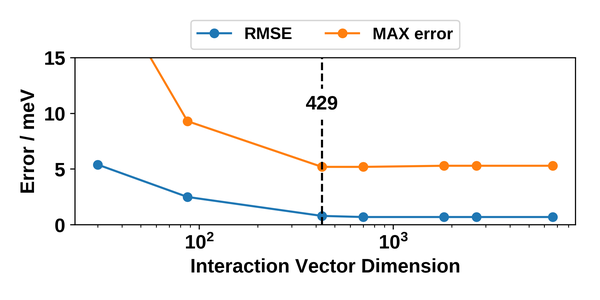}
	\caption{Convergence of the dimension of the {\interactionvector}}
	\label{fig:number_of_features}
\end{figure}
}

\section{DFT Convergence Tests}
\label{SI_sc:DFT_Convergence_Test}

\subsection{Lattice Constant}

\begin{figure}[h!]
\includegraphics[width=\linewidth]{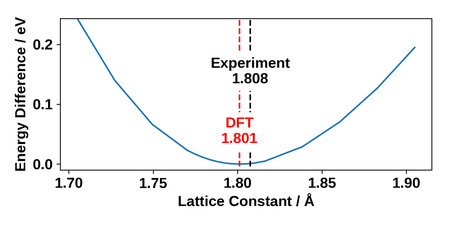}
\caption[Lattice constant convergence]{Lattice constant convergence}
\label{fig:lattice_constant}
\end{figure}

The lattice constant is converged for a primitive ffc-bulk unit cell with a k-grid containing $80$ k-points in each dimensions of the reciprocal unit cell (see figure \ref{fig:lattice_constant}). We find a lattice constant of $1.801~\angstrom$, which is slightly smaller than the experimental lattice constant $1.808~\angstrom$\cite{wyckoff1963interscience}.

\subsection{k-Grid}

\begin{figure}[h!]
  \includegraphics[width=\linewidth]{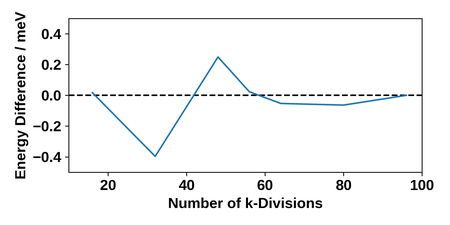}
\caption{Convergence of the k-gird}
\label{fig:k_point_convergence}
\end{figure}

The k-grid is converged for a primitive fcc-bulk unit cell with a lattice constant of $a = 3.61~\angstrom$ (see figure \ref{fig:k_point_convergence}). We use the densest k-grid setting as reference. We select a k-point density equivalent to $80 \times 80 \times 80$ k-points for a primitive unit cell, resulting in an uncertainty of about $0.1~meV$ per Cu atom. Since we use $5$ Cu layers for the substrate slab we get an uncertainty of $0.009~eV nm^{-2}$ ($0.1~meV$ per single copper atom or $0.5~meV/Cu$ per surface atom in the $5$ layer substrate).

\subsection{Number of Slab Layers}

\begin{figure}[h!]
	\includegraphics[width=\linewidth]{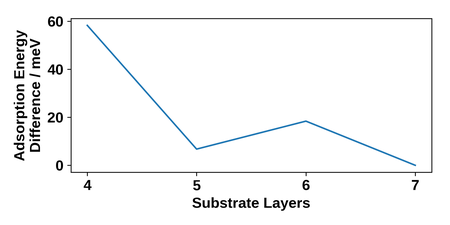}
	\caption{Layer convergence for slab}
	\label{fig:slab_layers}
\end{figure}

Using the repeated slab approach, we converge the number of substrate layers by considering the adsorption energy of a single naphthalene molecule in a $(4,0,0,4)$ unit cell (see figure \ref{fig:slab_layers}).  We use the seven layer slab as reference. The k-grid is set to $20$ k-points in both horizontal directions (along the primitive lattice vectors) of the reciprocal unit cell and one k-point in the z-direction (perpendicular to the surface). We find that $5$ Cu layers converge the uncertainty to about $0.018~eV nm^{-2}$ ($16~meV$ per unit cell with $16$ surface atoms).

\subsection{Onset of the cut-pot}

\begin{figure}[h!]
  \includegraphics[width=\linewidth]{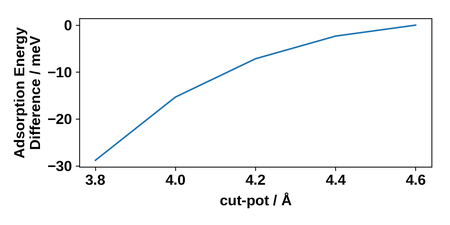}
\caption{Convergence of the cut-pot onset}
\label{fig:cutpot}
\end{figure}

The onset of the cutoff of the radial basis functions (first parameter of cut-pot) is converged using the adsorption energy of a single naphthalene molecule in a $(4,0,0,4)$ unit cell, with a k-grid containing $20$ k-points in the lateral and $1$ k-point in vertical directions (see figure \ref{fig:cutpot}). We use the largest cut-pot onset as reference. By setting the parameter cut-pot onset to $4.6~\angstrom$ the calculation time is increased by $+4~\%$, compared to the default setting. The remaining uncertainty is about $0.006~eV nm^{-2}$ ($5~meV$ per unit cell with $16$ surface atoms).

\subsection{Removal of the 5g Basis Function for Cu}

Removing the 5g basis function is tested using the adsorption energy of a single naphthalene molecule in a $(4,0,0,4)$ unit cell. The k-grid is set to $20$ k-points in the x- and y-direction of the reciprocal unit cell and one k-point in the z-direction. We find a reduction in calculation time of $54~\%$ and a resulting uncertainty of $0.017~eV nm^{-2}$ ($0.01543~eV$ per unit cell with $16$ surface atoms). Therefore, we use tier1 tight settings without the 5g basis function.

\begin{tabular}{lr}
Adsorption Energy (tier1 tight) & $-0.46273 ~eV$ \\
Adsorption Energy (tier1 tight without 5g) & $-0.47816 ~eV$ \\\hline
Difference & $0.01543 ~eV$
\end{tabular}

\subsection{Radial Multiplier}

Reducing the radial multiplier is tested using the adsorption energy of a single naphthalene molecule in a $(4,0,0,4)$ unit cell. The k-grid is set to $20$ k-points in the x- and y-direction of the reciprocal unit cell and one k-point in the z-direction. Reducing the radial multiplier to $1$ reduces the calculation time by $16~\%$ and results in an uncertainty of $0.0015~eV nm^{-2}$ ($0.00133 ~eV$ per unit cell with $16$ surface atoms).

\begin{tabular}{lr}
Adsorption Energy (Multiplier set to 2) & $-0.46273 ~eV$ \\
Adsorption Energy (Multiplier set to 1)  & $-0.46406 ~eV$ \\\hline
Difference & $0.00133 ~eV$
\end{tabular}

\subsection{Additional DFT Settings}

Apart from the aforementioned exceptions we use tier1 tight species settings. In addition, we use the following settings:

\begin{lstlisting}
# General Settings: 
xc	pbe
spin	none
charge	0
relativistic	atomic_zora scalar
occupation_type	gaussian 0.01
k_grid	20 20 1	 # for a (4,0,0,4) unit cell

# Convergence Criteria: 
sc_accuracy_forces	1e-3
sc_accuracy_rho	1e-2
sc_iter_limit	200
sc_accuracy_etot	1e-5

# Other Settings: 
collect_eigenvectors	.false.
RI_method	lvl_fast
vdw_correction_hirshfeld	.true.
compensate_multipole_errors	.true.
use_dipole_correction	.true.
\end{lstlisting}

\bibliographystyle{model1-num-names}
\bibliography{../references}